\documentclass[twocolumn,superscriptaddress,amsmath,amssymb,aps,physrev,floatfix]{revtex4-2}

\usepackage[colorlinks,plainpages=false,linkcolor=black,urlcolor=black,citecolor=black,pdfpagemode=UseNone,pdfstartview=FitBH]{hyperref}
\usepackage{graphicx}
\usepackage{dcolumn}
\usepackage{bm}
\usepackage{multirow}
\usepackage{subcaption}
\usepackage[justification=raggedright]{caption}

\usepackage{ulem}

\begin{document}

\title{\textbf{Magnetic phase diagram of magnetocaloric $\mathbf{TmFeO_3}$} 
}

\author{Kirill~I.~Tkachenko}
\affiliation{
Petersburg Nuclear Physics Institute by B.P.~Konstantinov of NRC ''Kurchatov Institute'', 188300 Gatchina, Russia
}

\author{Piotr~Fabrykiewicz}
\affiliation{
Institute of Crystallography, RWTH Aachen University, 52056 Aachen, Germany
}

\author{Aleksandr~K.~Ovsianikov}
\email{ovsianikov\_ak@pnpi.nrcki.ru}
\affiliation{
 Petersburg Nuclear Physics Institute by B.P.~Konstantinov of NRC ''Kurchatov Institute'', 188300 Gatchina, Russia
}

\author{Martin~Meven}
\affiliation{
Institute of Crystallography, RWTH Aachen University, 52056 Aachen, Germany
}

\author{Oleg~V.~Usmanov}
\affiliation{
 Petersburg Nuclear Physics Institute by B.P.~Konstantinov of NRC ''Kurchatov Institute'', 188300 Gatchina, Russia
}

\author{Igor~A.~Zobkalo}
\affiliation{
 Petersburg Nuclear Physics Institute by B.P.~Konstantinov of NRC ''Kurchatov Institute'', 188300 Gatchina, Russia
}

\author{Kirill~A.~Shaykhutdinov}
\affiliation{
 Kirensky Institute of Physics, Federal Research Center, Krasnoyarsk 660036, Russia
}
\affiliation{
 Siberian Federal University, Krasnoyarsk, 660071, Russia
}

\author{Konstantin~Yu.~Terentjev}
\affiliation{
 Petersburg Nuclear Physics Institute by B.P.~Konstantinov of NRC ''Kurchatov Institute'', 188300 Gatchina, Russia
}

\author{Sergey~V.~Semenov}
\affiliation{
 Kirensky Institute of Physics, Federal Research Center, Krasnoyarsk 660036, Russia
}
\affiliation{
 Siberian Federal University, Krasnoyarsk, 660071, Russia
}

\author{Eric~Ressouche}
\affiliation{Université Grenoble Alpes, CEA, IRIG, MEM, MDN, 38000 Grenoble, France
}

\author{Ketty~Beauvois}
\affiliation{Université Grenoble Alpes, CEA, IRIG, MEM, MDN, 38000 Grenoble, France
}

\date{\today}

\begin{abstract}
\noindent Neutron diffraction experiments of $\rm{TmFeO_3}$ single crystals were performed in the external magnetic fields. The field along \textit{c}-axis increases temperature of spin-reorientation transition $T_{SR}$ from phase $\Gamma 4$ to $\Gamma 2$. Application of the field along \textit{b}-axis led to the decrease of $T_{SR}$ and to the formation of new phases. Based on the temperature and field dependence of the Bragg reflection intensity, the configuration of magnetically induced phases was proposed.
\end{abstract}

\maketitle

\section{Introduction}
The rare earth orthoferrite family $\rm{RFeO_3}$, where R is a rare earth element, has attracted considerable attention for many years. The amazing magnetic properties of these materials are due to two magnetic subsystems -- iron and rare earths -- and their interactions. In addition to that, these compounds exhibit interesting and intriguing phenomena related to magnetism in them, such as magnetocaloric effects, multiferroicity, optical ultrafast manipulation of magnetic order, etc. The members of the family crystallize in an orthorhombic distorted perovskite structure with the space group \textit{Pnma} (also considered as \textit{Pbnm} in another setting). Orthoferrites $\rm{RFeO_3}$ are characterized by a rather high Néel temperatures $T_{N}$ ranging from 600~K to 740~K~\cite{White1969}. Below TN, the iron moments become antiferromagnetically ordered, with the sublattices being weakly canted due to the Dzyaloshinskii-Moriya interaction~\cite{Dzyaloshinsky1958, Moriya1960}, resulting in weak ferromagnetism. In the lower temperature range, the influence of the rare-earth -- iron exchange interaction leads to a spin reorientation (SR) transition, the temperature of which varies in a wide range for different R from 3 -- 10~K for $\rm{TbFeO_3}$~\cite{Artyukhin2012} up to 480~K for $\rm{SmFeO_3}$~\cite{Nikolov1994}.

Until recently, it was believed that ferroelectricity in orthoferrites is not allowed because it is theoretically prohibited by their centrosymmetric \textit{Pnma} structure. However, recent works suggest low-temperature ferroelectricity in $\rm{DyFeO_3}$~\cite{Tokunaga2008}, $\rm{GdFeO_3}$~\cite{Tokunaga2009}, $\rm{TbFeO_3}$~\cite{Song2014} and moreover, the observation of the emergence of room-temperature ferroelectricity in $\rm{SmFeO_3}$~\cite{Lee2011} and $\rm{YFeO_3}$~\cite{Shang2013}. This multiferroic behavior at higher temperatures opens up the possibility of using of such compounds for switching elements or sensors in the industry applications. Fast manipulation of the spin reorientation in $\rm{TmFeO_3}$ has been obtained by laser on ultrafast timescales~\cite{Kimel2004}. This could be a new direction for spintronics with high speed of magnetic recording. 

$\rm{TmFeO_3}$ crystallizes in the distorted perovskite structure, described by orthorhombic \textit{Pnma} space group. Below the Néel temperature $T_{N} = 635$~K~\cite{Bombik2003} $\rm{Fe^{3+}}$ ions order in $\Gamma 4$ phase with G-type~\cite{Bertaut1963} antiferromagnetic order along the \textit{c}-axis. Antisymmetric Dzyaloshinskii-Moriya interaction provides the canting of neighboring antiferromagnetically oriented spins, leading to a net magnetization \textbf{m} along the \textit{b}-axis. SR has been observed below $T_{SR} = 93$~K~\cite{Leake1968, Tsuyoshi1973}, and between 93 and 83~K, the crystal is in the mixed $\Gamma 24$ phase, with \textbf{m} laying in the \textit{bc} plane. Below 83~K the magnetic system is in $\Gamma 2$ phase, and \textbf{G}-type antiferromagnetic ordering  now directed along axis \textit{b}, while weak ferromagnetic component of type F directed along axis \textit{c}.

One of the most interesting effects associated with orthoferrites is the magnetocaloric effect, which represent itself as the change of temperature of a magnetic substance when subjected to an external magnetic field \textbf{B} under adiabatic conditions. For the studied compound, $\rm{TmFeO_3}$, the entropy change has a maximum $\Delta S \approx 12$~J/kg$\cdot$K at 17~K under an applied field of 7~T along the \textit{b} axis~\cite{Ke2016}. The appearance of $\Delta S$-peaks indirectly demonstrates the possibility of spin reorientation transitions in this field and temperature range in thulium orthoferrite. This requires the study of the phase diagram (\textit{H}, \textit{T}) of $\rm{TmFeO_3}$ in external magnetic fields. These types of diagrams help to provide a better understanding of the processes within the Tm subsystem. In addition, the external magnetic field will lead to changes in the energy balance and affect the exchange interaction and anisotropy within the thulium and iron subsystems. 

As it is known from a number of previous works on orthoferrites (e.g.~\cite{Skorobogatov2020, Ovsianikov2022}), the exchange interaction within the Fe-sublattice plays the main role in the formation of the magnetic structure below $T_N$. For $\rm{TmFeO_3}$ the strongest interaction is the superexchange one through O between the nearest neighbors in the Fe-chains along the \textit{b}-axis and within the \textit{ac}-plane $J_{b}^{\rm{Fe}} = 5.15$~meV and $J_{ac}^{\rm{Fe}} = 4.74$~meV respectively~\cite{Skorobogatov2020}.  The exchange field from ordered $\rm{Fe^{3+}}$ subsystem induces an ordered magnetic moment on $\rm{Tm^{3+}}$ ions, which is very weak at the temperature above $T_{SR}$, but becomes more noticeable below spin-reorientation transition~\cite{Fabrykiewicz2021}. Thus, when the temperature is decreased, the exchange between Fe and Tm sublattices, $J^{\rm{Fe-Tm}}$ becomes effective, which leads to spin reorientation transition.

\section{Experimental part}
 High quality single crystals of $\rm{TmFeO_3}$ were grown by the flux method~\cite{Barilo1991} using the optical floating zone technique (FZ-4000, Crystal Systems Corporation) with the natural isotope mixture. X-Ray studies were performed at the Rigaku SmartLab diffractometer at the Petersburg Nuclear Physics Institute (PNPI). These measurements were performed with wavelength 1.54~\AA, $2\theta$ range $20^{\circ} - 135^{\circ}$ and the temperature range $T = 50 - 300$~K using powder obtained by grinding of the crystal. Refinement of crystal structure was made using the FullProf Suite software package~\cite{RodriguezCarvajal}. Single crystal with cylindrical shape and dimensions of $4 \times 4 \times 8$~mm$^3$ was used for the neutron studies. The neutron diffraction experiments were performed at the Institut Laue-Langevin. The orientation of the crystal was obtained by OrientExpress~\cite{Ouladdiaf2006}. This is an automated Laue neutron diffractometer based upon two high-performance image-enhanced CCD cameras coupled to a large-area neutron scintillator and allows fast orientation of the crystal. Magnetic structure studies were performed on diffractometer D23~\cite{ILLD23} in external magnetic fields with wavelength 1.28~\AA. Two sets of measurements were made for the direction of magnetic fields along crystal axes \textit{b} and \textit{c} with constant field values. For both sample orientations, sets of selected peaks were measured for each combination at temperatures 2, 20, 40, 60, 75, 88, 100~K and magnetic fields 0, 0.5, 1, 2, 3, 4, 5, 5.8~T. In addition, for both sample orientations, we measured intensities of selected peaks at temperatures 80, 90, 95, 105, and 110~K without magnetic field; this temperature interval includes the temperature range of Fe spin reorientation. Furthermore, two sets of all available reflections (about 500) at temperature $T = 20$~K were collected for two different sample orientations. The data that support the findings of this article are openly available~\cite{Data}. The program Mag2Pol~\cite{JApplCryst2019} was used then to calculate and refine the magnetic structure.
 
\section{Result and discussion}
\subsection{Structure study}
 Our X-ray studies confirmed that the crystal structure of $\rm{TmFeO_3}$ is described by the space group \textit{Pnma} \#62~IT~\cite{InternationalTables2006} at all temperatures of measurement. In this structure Fe-ions occupy position 4\textit{b}, Tm -- 4\textit{c}, O1 and O2 -- 4\textit{c} and 8\textit{d} respectively. Obtained parameters of the crystal structure of $\rm{TmFeO_3}$ at different temperatures are shown in Table~\ref{tab:table1} and they agree well with those ones in works~\cite{Bombik2000}~\cite{Marezio1970}. 
The unit cell parameters increase monotonically with increasing temperature, that was also observed in ~\cite{Bombik2001, Khan2021}.
 
\begin{table} 
\caption{\label{tab:table1} Crystal structure of $\rm{TmFeO_3}$ at different temperatures, X-ray powder data, \textit{Pnma} setting.}
\begin{ruledtabular}
\begin{tabular}{cccccc}
 \multicolumn{2}{l}{$\rm{TmFeO_3}$} & 300~K & 150~K & 85~K & 50~K \\
 \hline
 $a$ [\AA] & & 5.5721(2) & 5.5679(2) & 5.5655(2) & 5.5650(2) \\
 $b$ [\AA] & & 7.5823(2) & 7.5761(3) & 7.5740(3) & 7.5732(3) \\
 $c$ [\AA] & & 5.2468(2) & 5.2429(2) & 5.2415(2) & 5.2407(2) \\
 \hline
 \multirow{3}{*}{Fe (4\textit{b})} & $x$ & 0.5000 & 0.5000 & 0.5000 & 0.5000 \\
  & $y$ & 0.0000 & 0.0000 & 0.0000 & 0.0000 \\
  & $z$ & 0.0000 & 0.0000 & 0.0000 & 0.0000 \\
 \hline
 \multirow{3}{*}{Tm (4\textit{c})} & $x$ & 0.0690(2) & 0.0693(2) & 0.0693(2) & 0.0691(2) \\
  & $y$ & 0.2500 & 0.2500 & 0.2500 & 0.2500 \\
  & $z$ & 0.9839(3) & 0.9839(3) & 0.9831(3) & 0.9820(3) \\
 \hline
 \multirow{3}{*}{O1 (4\textit{c})} & $x$ & 0.455(2) & 0.459(2) & 0.457(2) & 0.459(2) \\
  & $y$ & 0.2500 & 0.2500 & 0.2500 & 0.2500 \\
  & $z$ & 0.114(2) & 0.115(2) & 0.111(2) & 0.117(2) \\
 \hline
 \multirow{3}{*}{O2 (8\textit{d})} & $x$ & 0.806(2) & 0.803(2) & 0.803(2) & 0.807(2) \\
  & $y$ & 0.058(1) & 0.060(1) & 0.058(1) & 0.058(1) \\
  & $z$ & 0.807(1) & 0.804(2) & 0.806(1) & 0.806(1) \\
 \hline
  & & $\chi^2 = 1.15$ & $\chi^2 = 1.27$ & $\chi^2 = 1.13$ & $\chi^2 = 1.05$
\end{tabular}
\end{ruledtabular}
\end{table}

\begin{figure*} 
\includegraphics[width=\linewidth]{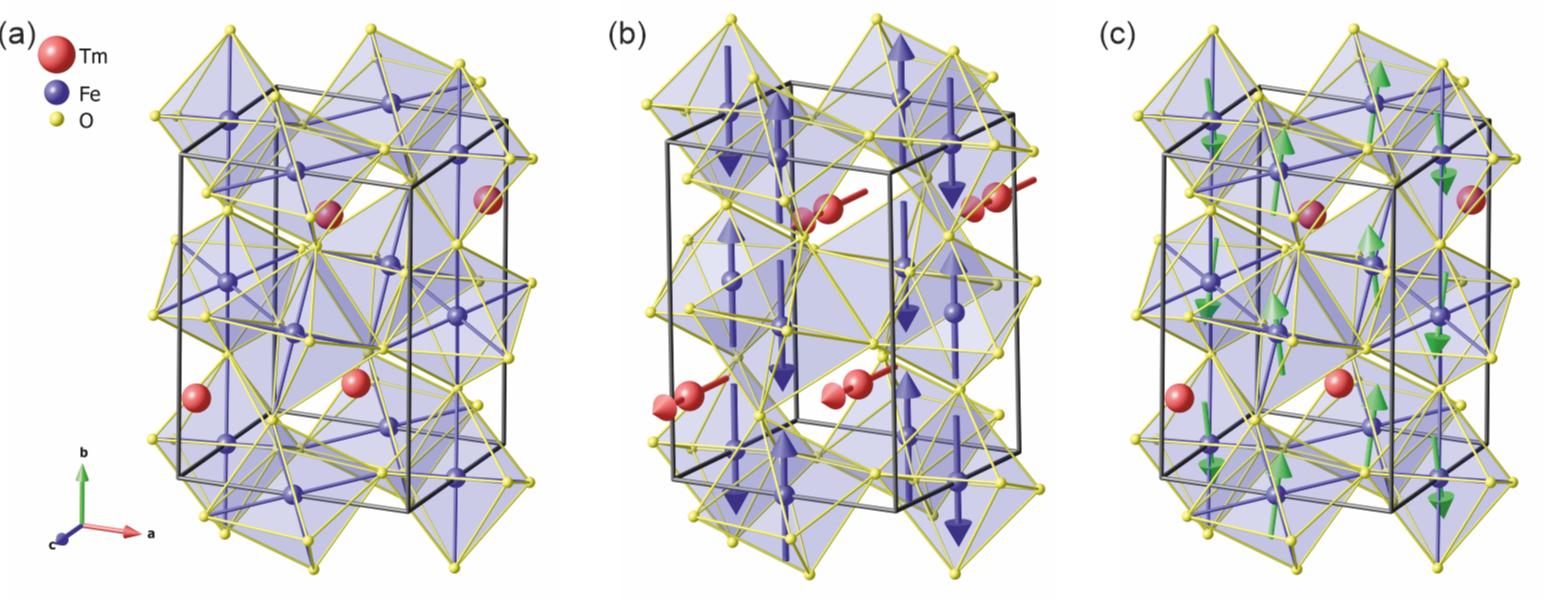}
\caption{\label{fig:fig1} (a) Crystal structure of $\rm{TmFeO_3}$ at $T=300$~K; (b) Magnetic structure of $\rm{TmFeO_3}$ at $T=20$~K. Blue arrows -- Fe magnetic moments, red arrows -- Tm magnetic moments, the magnitude of the arrow for Tm is increased five times for the sake of visualization; c) Antiferroelectric order of $\rm{TmFeO_3}$, green arrows -- electric dipole moment.}
\end{figure*}

$\rm{Fe^{3+}}$ ions occupy the 4\textit{b} position in the \textit{Pnma} unit cell (see Fig.~\ref{fig:fig1}a) and there are four types of possible collinear ordering of the Fe subsystem, which can be expressed by means of the following Bertaut notation~\cite{Bertaut1963}:

\begin{eqnarray} 
\mathbf{F = S_1 + S_2 + S_3 + S_4}, \nonumber \\
\mathbf{C = S_1 + S_2 - S_3 - S_4}, \nonumber \\
\mathbf{A = S_1 - S_2 - S_3 + S_4}, \nonumber \\
\mathbf{G = S_1 - S_2 + S_3 - S_4}, \nonumber
\end{eqnarray}

\noindent where \textbf{G} describes the main antiferromagnetic component of the magnetic structure, \textbf{F} is the ferromagnetic vector, weak antiferromagnetic components \textbf{C} and \textbf{A} describe the weak canting of the magnetic moments. It is convenient situation to detect the mode of magnetic alignment of elements at 4\textit{b} site in the crystal with space group \textit{Pnma}. In this case different magnetic modes give contributions to Bragg reflections with different parity: mode \textbf{A} gives a contribution to reflections with $h + l$ even, $k$ odd; \textbf{C} --  $k$ even, $h + l$ odd; \textbf{F} -- $h + l$ even, $k$ even; \textbf{G} -- $h + l$ odd, $k$ odd (see Table~\ref{tab:table2}). $\mathbf{S_1, S_2, S_3, S_4}$ means projections of Fe magnetic moments on arbitrary axis.  In case $\rm{TmFeO_3}$, Fe-magnetic sublattice is ordered below temperature $T_{N} = 635$~K in a G-type structure $\Gamma 4$ ($A_x, F_y, G_z$). Irreducible representation $\Gamma 4$ of the space group \textit{Pnma} is build based on the propagation vector $\textbf{k} = (0, 0, 0)$ and it is equivalent to \textit{Pn’ma’} magnetic space group (see Table~\ref{tab:table3}) with small spin canting caused by the Dzyaloshinskii-Moriya (DM) interaction that induces a weak ferromagnetic (FM) moment. Spin-reorientation transition begins at decreasing temperature below $T_{SR1} = 93$~K. This transition goes through the mixed phase $\Gamma 24$ and at temperature $T_{SR2} = 83$~K the main G-type component of Fe-spins lies along the b-axis and the magnetic phase $\Gamma 2 \ (C_x, G_y, F_z)$ is forming in the compound. Symmetry analysis~\cite{Przenioslo2018} shows that continuous spin reorientation requires symmetry lowering at least to a monoclinic one. This statement is true only for continuous spin reorientation (i.e. for one magnetic domain where magnetic moment continuously changes direction with continuous change of temperature), while spin reorientation in $\rm{TmFeO_3}$ has been reported in the form of changing the population of magnetic domains~\cite{Tsymbal2010} -- such mechanism of spin reorientation is compatible with orthorhombic symmetry. In our studies, we have not observed lowering orthorhombic symmetry of lattice in the spin reorientation region that is also in the agreement with~\cite{Tsymbal2006, Tsymbal2007}. As discussed in~\cite{Fabrykiewicz2021, Fabrykiewicz2023} both modes below and above the spin reorientation transition correspond to the same 'AFB' set of directions of magnetic moment. Detailed crystallographic and magnetic symmetry analysis of rare-earth orthoferrites was presented in~\cite{Fabrykiewicz2021} and analysis extension to (anti)ferroelectric and toroidal moments ordering was discussed in~\cite{Fabrykiewicz2023}. 

The magnetic structure was refined simultaneously for two datasets collected with different crystal orientations at $T = 20$~K without magnetic field. In magnetic neutron scattering, the magnetic form factor $F(q)$ defines the $q$-dependence of the magnetic scattering amplitude of a single ion in a such way that the magnetic signal at large $q$ -- large angles -- is negligible in comparison with the nuclear one. Since we have no dataset above Néel temperature, in the first stage of fitting, reflections measured at $|Q| > 8.5$~r.l.u. (two theta angles greater than 60 degrees) were selected for the structure refinement from the collected dataset and the scale factor and extinction parameters were fitted. In the second stage of fitting, the remaining reflections, collected at low angles, were used to estimate the values of Fe and Tm magnetic moments. Neutron paths in the sample were close each other in magnitude for reflections with different $hkl$, therefore absorption correction did not affect a lot on the results of the refinement. Then magnetic moments values were obtained to be equal $M_x^{Fe}=0$, $M_y^{Fe}=4.93(22)$, $M_z^{Fe}=-0.51(9)$ and $M_x^{Tm}=0$, $M_y^{Tm}=0$, $M_z^{Tm}=0.37(8)$ $\mu_B$ with R-factor = 11.2. The refinement with free parameters for $M_x^{Fe}$ gives zero for these values within esds and $M_y^{Tm}$ zero due to the symmetry restrictions, then these parameters were fixed equal to zero. Magnetic ordering of the $\rm{Tm^{3+}}$ sublattice corresponds to the irreducible representation $\Gamma 2$. The obtained magnetic moments values are in good agreement with those ones, obtained for $\rm{TmFeO_3}$ in work~\cite{Staub2017}. Figure~\ref{fig:fig1}b shows the obtained magnetic structure at $T = 20$~K without magnetic field. 

At the same time, X-ray measurements show that $\rm{Fe^{3+}}$ ions are not shifted from position 4\textit{b} above as well as below spin-reorientation transition within \textit{Pnma} group. The treatment with lower symmetry groups did not improve the crystal structure refinement and was skipped out therefore in the ongoing discussion. This allows us to choose $\rm{Fe^{3+}}$ as the center of the local coordinate system for determining the electric dipole moment. For that we apply the approach that was first proposed by A.~K.~Zvezdin in work~\cite{Zvezdin2021} and tested on rare-earth orthochromites. As it was shown in~\cite{Zvezdin2021}, in this structure electric dipole moments can appear due to the deviation of $\rm{O}^{2-}$ ions from their high symmetry positions in the parent perovskite structure. The basic ferroelectric modes can be described in a way analogous to the magnetic modes~\cite{Zvezdin2021}:

\begin{eqnarray} 
\mathbf{P = d_1 + d_2 + d_3 + d_4}, \nonumber \\
\mathbf{D = d_1 + d_2 - d_3 - d_4}, \nonumber \\
\mathbf{Q_2 = d_1 - d_2 - d_3 + d_4}, \nonumber \\
\mathbf{Q_3 = d_1 - d_2 + d_3 - d_4}, \nonumber
\end{eqnarray} 

\noindent where $\textbf{d}_i$ -- electric dipole moments in the vicinity of four $\rm{Fe^{3+}}$ ions; $\mathbf{P}$, $\mathbf{D}$, $\mathbf{Q_2}$, $\mathbf{Q_3}$ -- combinations of vectors transformed by the symmetry elements of space group \textit{Pnma}. Following this approach, the electric dipole moment for $\rm{TmFeO_3}$ at temperatures above and below spin-reorientation transition was obtained using X-ray data. It is shown that while the magnetic moment of Fe rotates during the transition of the magnetic structure from the $\Gamma 4$  to phase $\Gamma 2$, the electric dipole moment ordering did not change. The orientation vector of the electric dipole moment has coordinates $\mathbf{r}_{EDM} = (-0.09(1), -0.98(1), -0.06(1))$ and electric dipoles ordered in antiferroelectric D mode. Figure~\ref{fig:fig1}c shows the arrangement of electric dipole moments in $\rm{TmFeO_3}$ unit cell.

\begin{table}[h] 
\caption{\label{tab:table2} Correspondence of (anti)ferroelectric and magnetic modes in \textit{Pnma} crystal $\rm{TmFeO_3}$. Lower part of the Table show which ($hkl$) reflections gives intensity of each magnetic mode. ''\textit{e}'' and ''\textit{o}'' symbols denote even or odd parity of Miller indices, respectively.}
\begin{ruledtabular}
\begin{tabular}{ccllcccc}
  & & \multirow{2}{*}{Fe (4\textit{b})}& \multirow{2}{*}{Tm (4\textit{c})} & P & D & Q$_2$ & Q$_3$ \\
  & & & & F & A & C & G \\
 \hline
$\mathbf{d_1}$& $\mathbf{S_1}$ &
  $\left( 0, 0, \frac{1}{2} \right)$
  & $\left( x, \frac{1}{4}, z \right)$
  & + & + & + & + \\[2pt]

$\mathbf{d_2}$& $\mathbf{S_2}$ &
 $\left( 0, \frac{1}{2}, \frac{1}{2} \right)$
  & $\left( -x, \frac{3}{4}, -z \right)$
  & + & $-$ & + & $-$ \\[2pt]

$\mathbf{d_3}$& $\mathbf{S_3}$ &
 $\left( \frac{1}{2}, \frac{1}{2}, 0 \right)$
  & $\left( -x+\frac{1}{2}, \frac{3}{4}, z+\frac{1}{2} \right)$
  & + & $-$ & $-$ & + \\[2pt]

$\mathbf{d_4}$& $\mathbf{S_4}$ &
 $\left( \frac{1}{2}, 0, 0 \right)$
  & $\left( x+\frac{1}{2}, \frac{1}{4}, -z+\frac{1}{2} \right)$
  & + & + & $-$ & $-$ \\[1pt]
 \hline
& &
  & \multicolumn{1}{r}{$h+l$} & \textit{e} & \textit{e} & \textit{o} & \textit{o} \\
& &
  & \multicolumn{1}{r}{$k$} & \textit{e} & \textit{o} & \textit{e} & \textit{o} \\
\end{tabular}
\end{ruledtabular}
\end{table} 

\begin{table} 
\caption{\label{tab:table3} Irreducible representations (irreps) of the space group \textit{Pnma} built based on the $\mathbf{k}$-vector equal to ($0, 0, 0$) with corresponding magnetic space groups (MSG) and allowed magnetic modes for 4\textit{b} and 4\textit{c} Wyckoff positions.}
\begin{ruledtabular}
    \begin{tabular}{llcc}
        Irrep & MSG & Fe (4\textit{b}) & Tm (4\textit{c}) \\
        \hline
        $\Gamma 1$ & \textit{Pnma.1}		& ($G_x, C_y, A_z$) & ($0_x, C_y, 0_z$) \\
        $\Gamma 2$ & \textit{Pn’m’a}	& ($C_x, G_y, F_z$) & ($C_x, 0_y, F_z$) \\
        $\Gamma 3$ & \textit{Pnm’a’}	& ($F_x, A_y, C_z$) & ($F_x, 0_y, C_z$) \\
        $\Gamma 4$ & \textit{Pn’ma’}	& ($A_x, F_y, G_z$) & ($0_x, F_y, 0_z$) \\
        $\Gamma 5$ & \textit{Pn’m’a’}	& ($0_x, 0_y, 0_z$) & ($A_x, 0_y, G_z$) \\
        $\Gamma 6$ & \textit{Pnma’}		& ($0_x, 0_y, 0_z$) & ($0_x, A_y, 0_z$) \\
        $\Gamma 7$ & \textit{Pn’ma}		& ($0_x, 0_y, 0_z$) & ($0_x, G_y, 0_z$) \\
        $\Gamma 8$ & \textit{Pnm’a}		& ($0_x, 0_y, 0_z$) & ($G_x, 0_y, A_z$)
    \end{tabular}
\end{ruledtabular}
\end{table}

\subsection{Magnetic scattering dependence on magnetic field}
In order to change the balance of exchange and anisotropy interactions it is possible to add an additional energy parameter to the system, such as an external magnetic field. This could lead to a change in the spin-reorientation transition temperature as well as to a change in the magnetic structure. Since it is convenient to follow the changes in the magnetic alignment of $\rm{Fe^{3+}}$ ions by measuring Bragg reflections with definite parity, a few reflections were chosen and were measured in the temperature range of 2 -- 100~K without and with the external magnetic field applied along the \textit{b}- and \textit{c}-axes. 
 
\begin{figure*}[h!] 
  \centering
  \begin{subfigure}[b]{0.45\linewidth}
    \includegraphics[width=\linewidth]{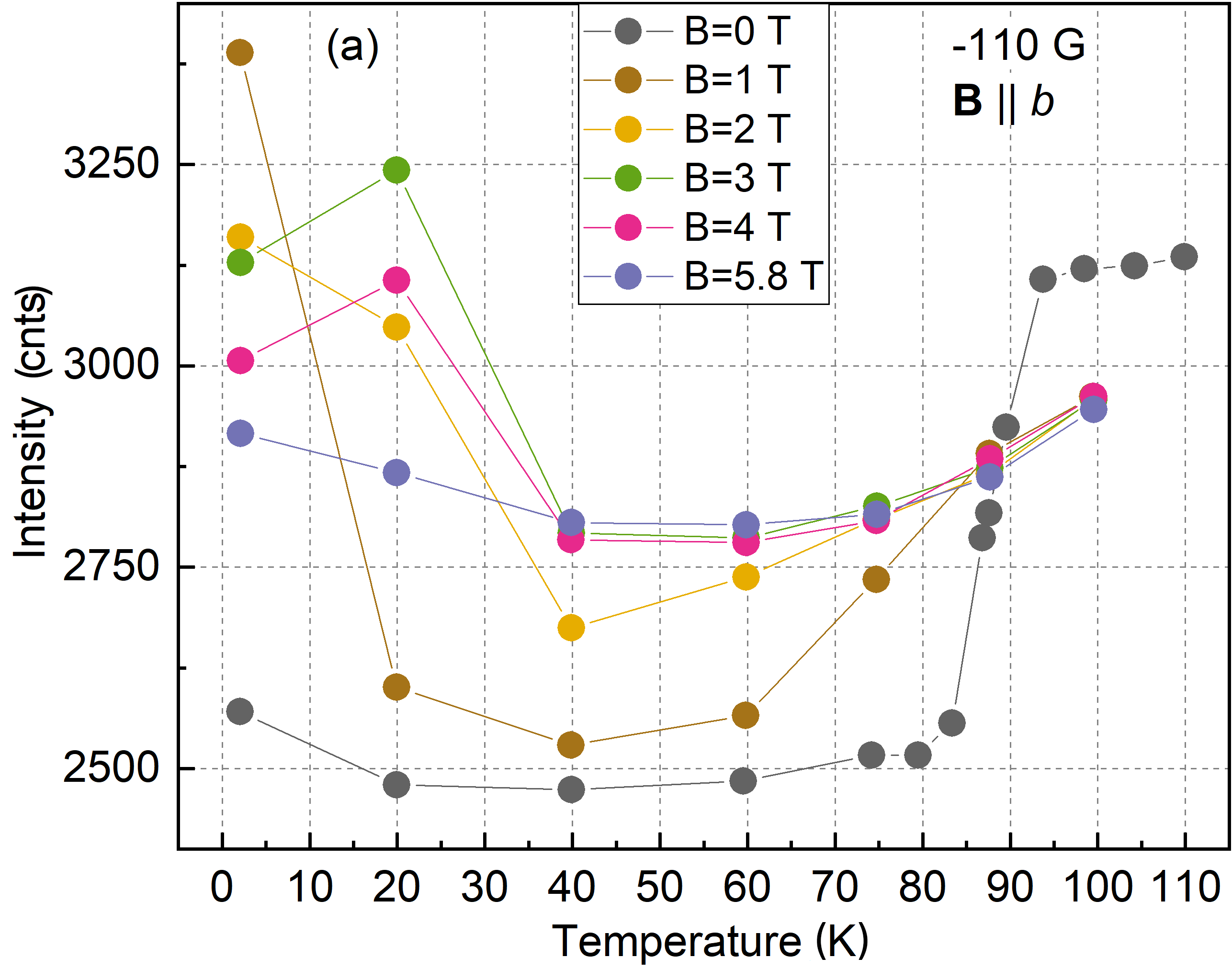}
  \end{subfigure}
  \begin{subfigure}[b]{0.45\linewidth}
    \includegraphics[width=\linewidth]{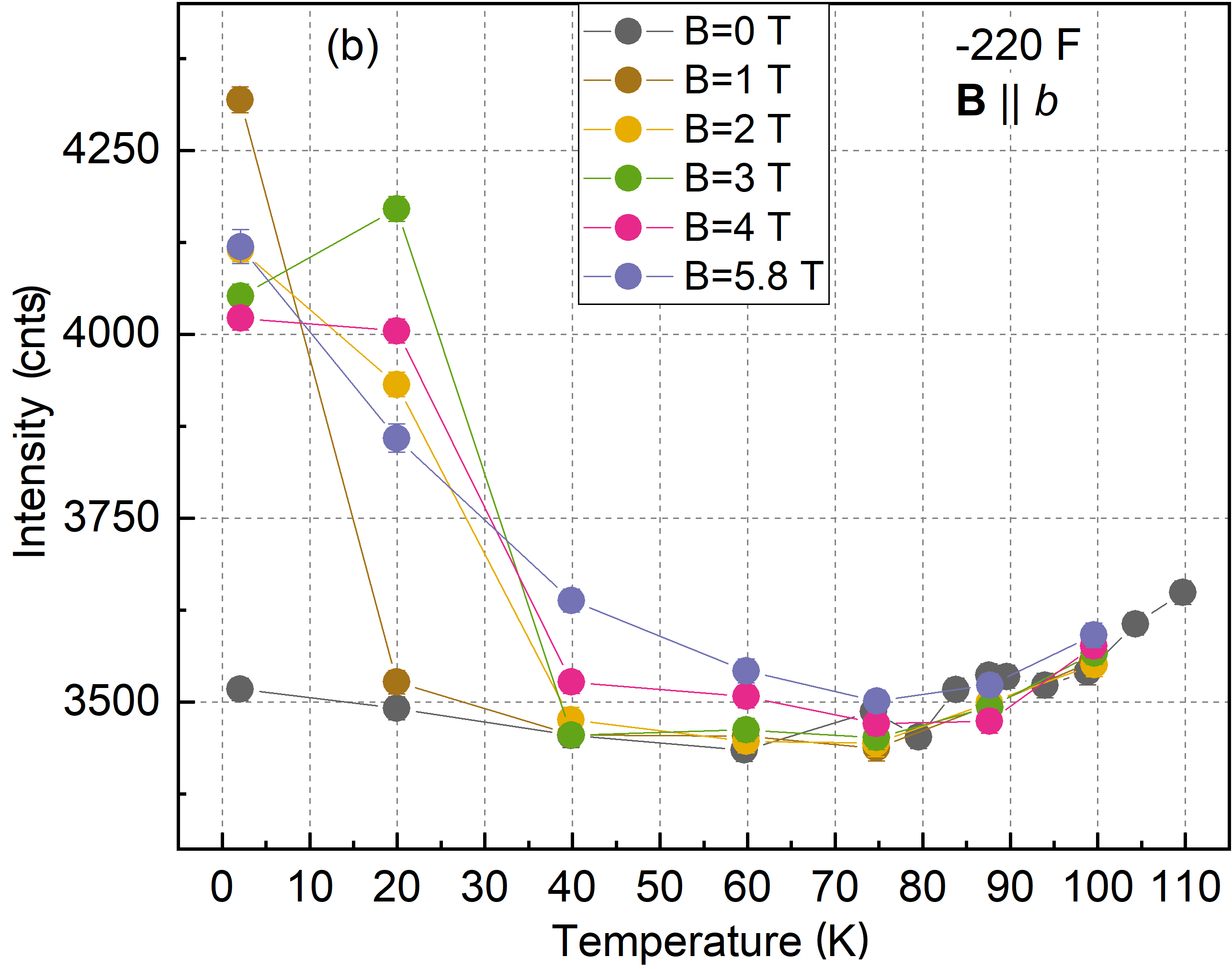}
  \end{subfigure}
  \begin{subfigure}[b]{0.45\linewidth}
    \includegraphics[width=\linewidth]{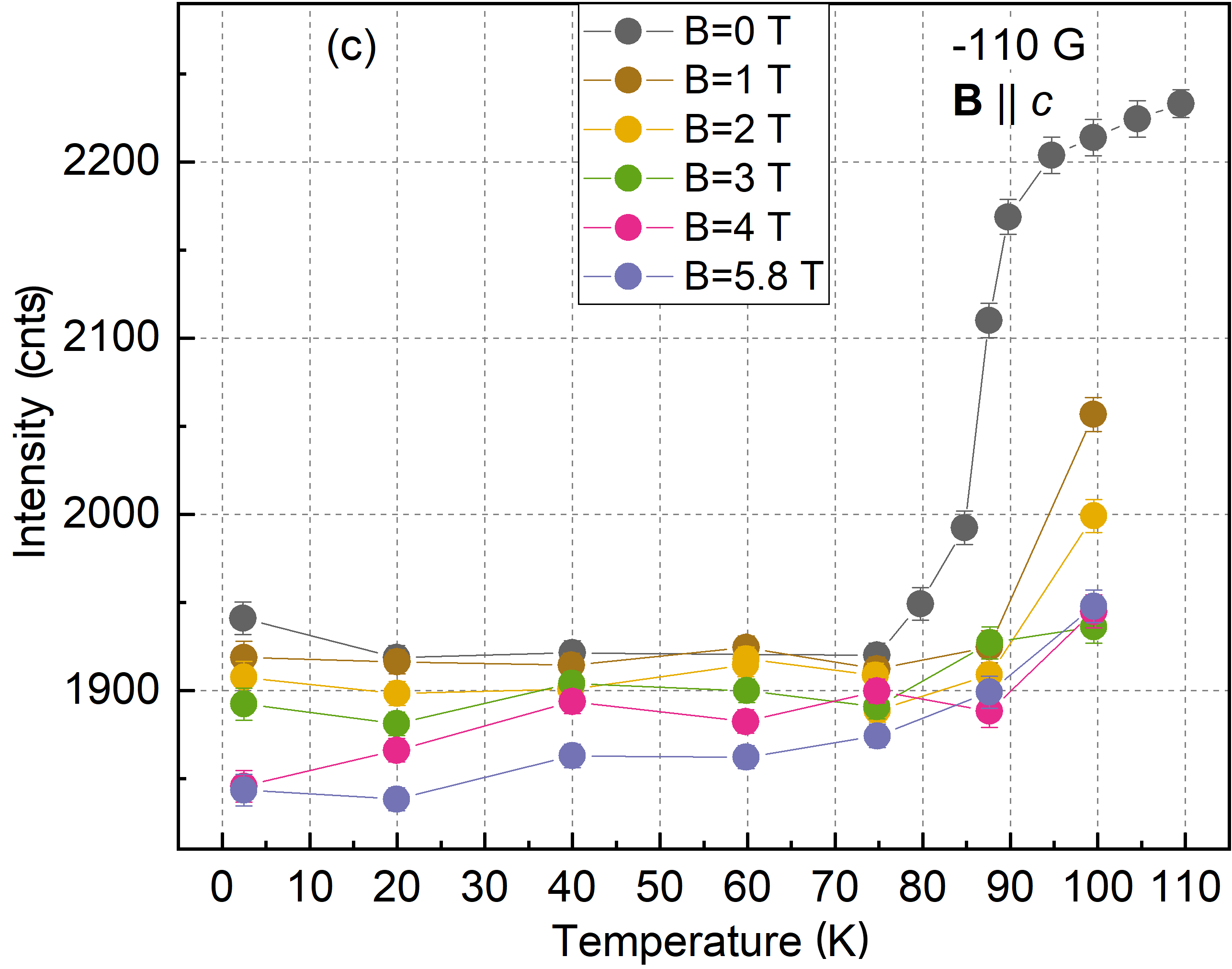}
  \end{subfigure}
  \begin{subfigure}[b]{0.45\linewidth}
        \includegraphics[width=\linewidth]{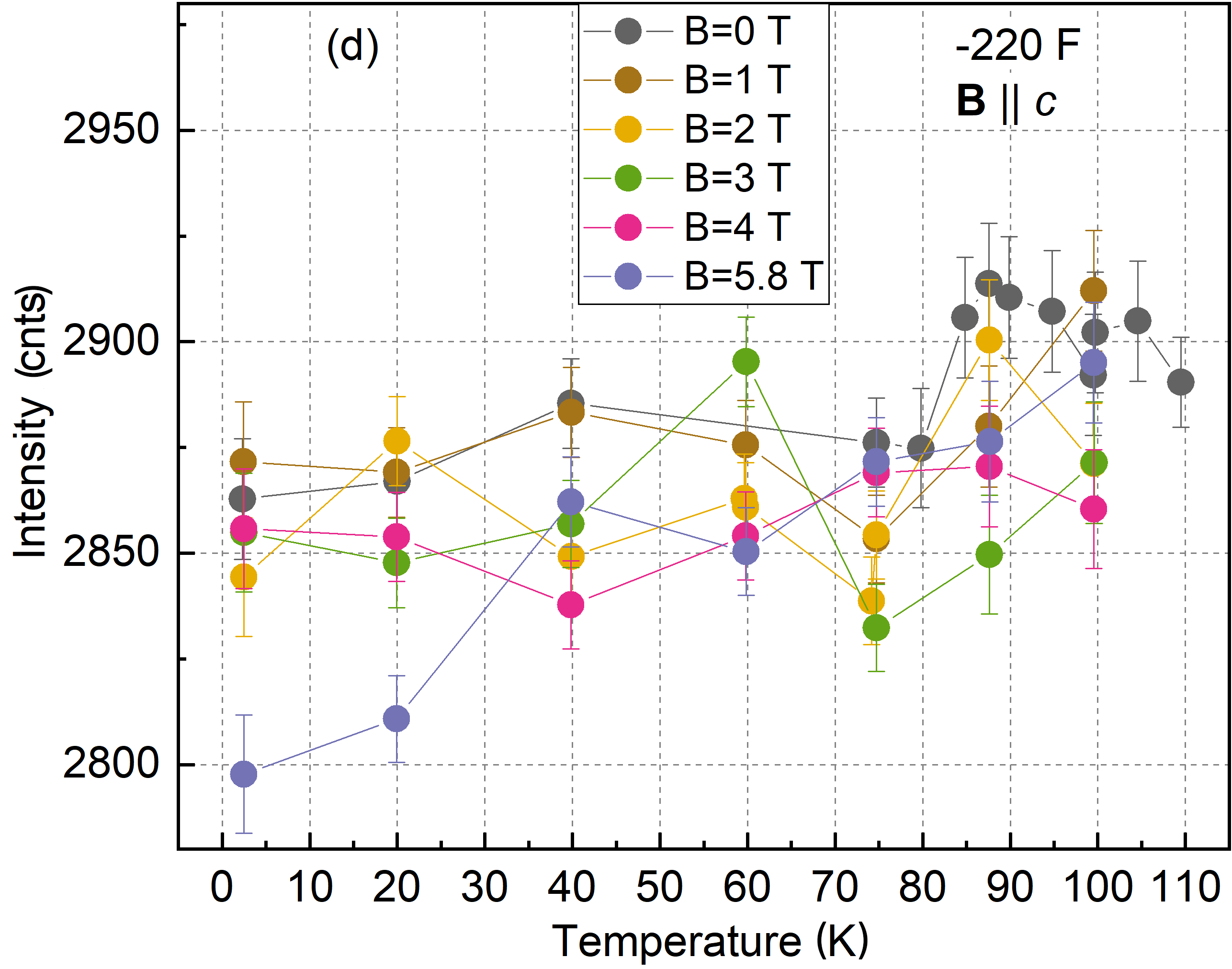}
  \end{subfigure}
  \caption{\label{fig:fig2} Temperature dependencies of reflections intensities ($-110$), ($-220$) for external magnetic field directed a) b) along the \textit{b}-axis and c) d) along the \textit{c}-axis.}
\end{figure*}

Figure~\ref{fig:fig2} shows temperature dependencies two of them: reflections $(-110)$ (G-type) and $(-220)$ (F-type). The intensity fall at the zero field dependence of the G-type reflection in the temperature range of 83 -- 93~K (Figure~\ref{fig:fig2}a) corresponds to the spin-reorientation transition from $\Gamma 4$ to $\Gamma 2$ phase through mixed phase $\Gamma 24$ with the rotation of the Fe main antiferromagnetic G-type order from the \textit{c}-axis to the \textit{b}-axis. The effective magnetic moment in that case decreases about twice, that leads to such intensity changes. The small increase in intensity below $T \approx 40$~K is probably associated with the additional insert of the induced magnetic moment on $\rm{Tm^{3+}}$ ions. The F-type reflection $(-220)$ keeps its intensity at zero field temperature dependence as expected. 

The application of a magnetic field in two different directions has a different effect on the temperature dependence of Bragg reflections. At the field along \textit{b}-axis, intensity of G-type reflection $(-110)$ goes to lower values in much smoother manner then it was without field, reaching lowest values at lower temperatures. It can also be seen that these lowest values are higher than that one which was observed without the field. And, at some temperatures -- below 20~K for 1~T, below 40~K for 2 -- 4~T -- the intensity of $(-110)$ increases considerably. The similar situation takes place for the F-type reflection $(-220)$ that is increase of intensity at low temperatures when field was applied along \textit{b}-axis (Figure~\ref{fig:fig2}b). The temperature behavior of these reflections at field 5.8~T looks very smooth and its intensity for $(-110)$ does not reach such a high value, as it was observed for smaller fields. 

Quite another situation was observed for the case when external magnetic field was directed along \textit{c}-axis. For that field direction intensity of G-type reflection goes to minimum value at a little higher temperatures than without field. In addition, no rise of intensity was observed when temperature goes down at any field value (Figure~\ref{fig:fig2}c). Regarding the temperature evolution of the F-type reflection for a given field direction, it seems to be possible to estimate that there is almost no influence of the magnetic field (Figure~\ref{fig:fig2}d).

We connect this situation with participation of rare earth magnetic subsystem. The $\rm{Tm^{3+}}$ ion is a non-Kramers' one with two singlets in ground states and it has a highly anisotropic \textit{g}-factor $g_x < 0.25$, $g_y = 8.1$ and $g_z < 0.25$~\cite{Malozemoff1970}. The orientation of the local coordinate system of $\rm{Tm^{3+}}$ ions is governed by the crystal field, and directly related to the tilts of oxygen octahedra. Thus, an external magnetic field directed along the \textit{b}-axis induces order on magnetic moments on the $\rm{Tm^{3+}}$ ions. For the case when field is along \textit{c}-axis, $\rm{Tm^{3+}}$ local susceptibility in this direction is negligible; therefore, no additional scattering is observed at Figure~\ref{fig:fig2}c,~d.  We suppose that temperature dependence of G- and F-type reflections at field 5.8~T could be related with the flop of the rare earth magnetic subsystem that is with so-called type-II switching. In type-II spin switching at a certain temperature point only moments of $\rm{R^{3+}}$ reverse thus turning parallel arrangement of $\rm{Fe^{3+}}$ and $\rm{R^{3+}}$ moments into antiparallel one or vice versa -- turning antiparallel one to parallel one. Spin switching of the second type is rarely found in the $\rm{RFeO_3}$ family. More common phenomenon is type-I spin switching where both $\rm{Fe^{3+}}$ and $\rm{R^{3+}}$ moments (induced due to superexchange interaction between $\rm{Fe^{3+}}$ and $\rm{R^{3+}}$) are simultaneously inverted at specific temperature point. At the same time, there is report in work~\cite{Song2023} about temperature-induced type-II spin switching in $\rm{TmFeO_3}$, thus one can suppose that in higher fields magnetic system of $\rm{Tm^{3+}}$ is in counterphase to $\rm{Fe^{3+}}$ one, thus producing near compensating situation.

At the low temperature and high magnetic fields we have also observed significant increase of intensity of A-type peaks ($-410$), ($-210$), ($210$), moderate increase of intensity of ($-111$) and small/negligible change of intensity of ($113$), ($012$) reflections. These changes are connected with the contribution from rare earth subsystem, as could be seen from following refinements. This behavior is indirectly correlated with the data of magnetic entropy change $\Delta S$, where the maximum of $\Delta S$ lies at 10~K in fields up to 3~T, and then shifts to the 15 -- 20~K temperature range at fields over 3~T~\cite{Ke2016}.

The sets of some tens reflections collected at temperatures 2, 20, 40, 60, 75, 88, 100~K and magnetic fields along 0, 0.5, 1, 2, 3, 4, 5, 5.8~T permitted to perform the refinement of magnetic structures realized at different condition (see Table~\ref{tab:table4}). The accuracy of the magnetic moment determination is defined by the comparatively low number of measured peaks. One could mark, that in the $\Gamma 4$ phase magnetic moments on $\rm{Fe^{3+}}$ ions do not change their values under the applied field keeping $M_z^{Fe}=4.7(2) \ \mu_B$ for $G_z$ and $M_y^{Fe} = 0.3(1)\ \mu_B$ for $F_y$. While $F_y$-component of $Tm^{3+}$ ions in $\Gamma 4$ demonstrate linear dependence on the applied field, thus changing from $M_y^{Tm} = 0.5(2)\ \mu_B$ at zero field up to $M_y^{Tm} = 1.3(3)\ \mu_B$ at 5.8~T field along \textit{b}-axis. A different situation for that field direction takes place in $\Gamma 2$ phase. In this phase all magnetic components keep their values independent on external field. The refinement of the mixed phase $\Gamma 24$ provides confirmation, that this phase consists of the mixture of two magnetic domains with configuration $\Gamma 4$ and $\Gamma 2$ that is consistent with orthorhombic symmetry. For the temperature range 2 -- 30~K and fields 2 -- 4~T along \textit{b}-axis, the refinement gives for $\rm{Fe^{3+}}$ subsystem antiferromagnetic alignment along \textit{c}-axis of type $G_z$ and ferromagnetic alignment of type, that is characteristic of $\Gamma 4$ configuration. At the same time, for $\rm{Tm^{3+}}$ subsystem, the refinement gives non-zero $C_x$ and $F_z$ components, that corresponds to $\Gamma 2$ configuration. For magnetic fields 5 -- 5.8~T at the temperatures 2 -- 30~K we obtained $F_y$ and $G_z$ components for $\rm{Fe^{3+}}$, and non-zero $F_y$-component for $\rm{Tm^{3+}}$ subsystem, which depending from magnetic field has values $M_y^{Tm} = -0.5(3) - -1.0(5)\ \mu_B$.

\begin{table}[h] 
\caption{\label{tab:table4} 
Magnetic structures realized in $\rm{TmFeO_3}$ at different temperatures and magnetic fields. Roman numbers in column ''Configuration'' correspond to spaces at phase diagram (Fig.~\ref{fig:fig3}). The specified values of magnetic moments are the average values for each configuration.}
\begin{ruledtabular}
\begin{tabular}{ccccc}
Configuration & Irrep & Mode & $M_{Fe}, \ \mu_B$ & $M_{Tm}, \ \mu_B$ \\ \hline
\multirow{3}{*}{I} & \multirow{3}{*}{$\Gamma 4$} & $A_x$ &  & \\
& & $F_y$ & 0.3(1) & 0.5(2) -- 1.3(2) \\
& & $G_z$ & 4.7(2) & \\ 
\hline
\multirow{6}{*}{II} & \multirow{3}{*}{$\Gamma 2$} & $C_x$ &  & \\
& & $G_y$ & 4.9(2) & \\
& & $F_z$ & -0.4(1) & 0.4(1) \\ 
\cline{2-5}
& \multirow{3}{*}{$\Gamma 4$} & $A_x$ &  & \\
& & $F_y$ & 0.3(1) & 0.5 -- 1.3(2) \\
& & $G_z$ & 4.8(2) & \\ 
\hline
\multirow{3}{*}{III} & \multirow{3}{*}{$\Gamma 2$}  & $C_x$ &  &  \\
& & $G_y$ & 4.9(2) & \\
& & $F_z$ & -0.4(1) & 0.4(1) \\
\hline
\multirow{6}{*}{IV} & \multirow{3}{*}{$\Gamma^{Fe}_4$} & $A_x$ &  & \\
& & $F_y$ & 0.3(1) & \\
& & $G_z$ & 4.8(3) & \\ 
\cline{2-5}
& \multirow{3}{*}{$\Gamma^{Tm}_2$} & $C_x$ &  & 0.3(1) \\
& & $G_y$ &  &  \\
& & $F_z$ &  & 0.4(1) \\ 
\hline
\multirow{3}{*}{V} & \multirow{3}{*}{$\Gamma 4$}  & $A_x$ &  &  \\
& & $F_y$ & 0.3(1) & -0.5(3) -- -1.0(3) \\
& & $G_z$ & 4.7(3) & 
\end{tabular}
\end{ruledtabular}
\end{table}

\begin{figure} 
\includegraphics[width=0.45\textwidth]{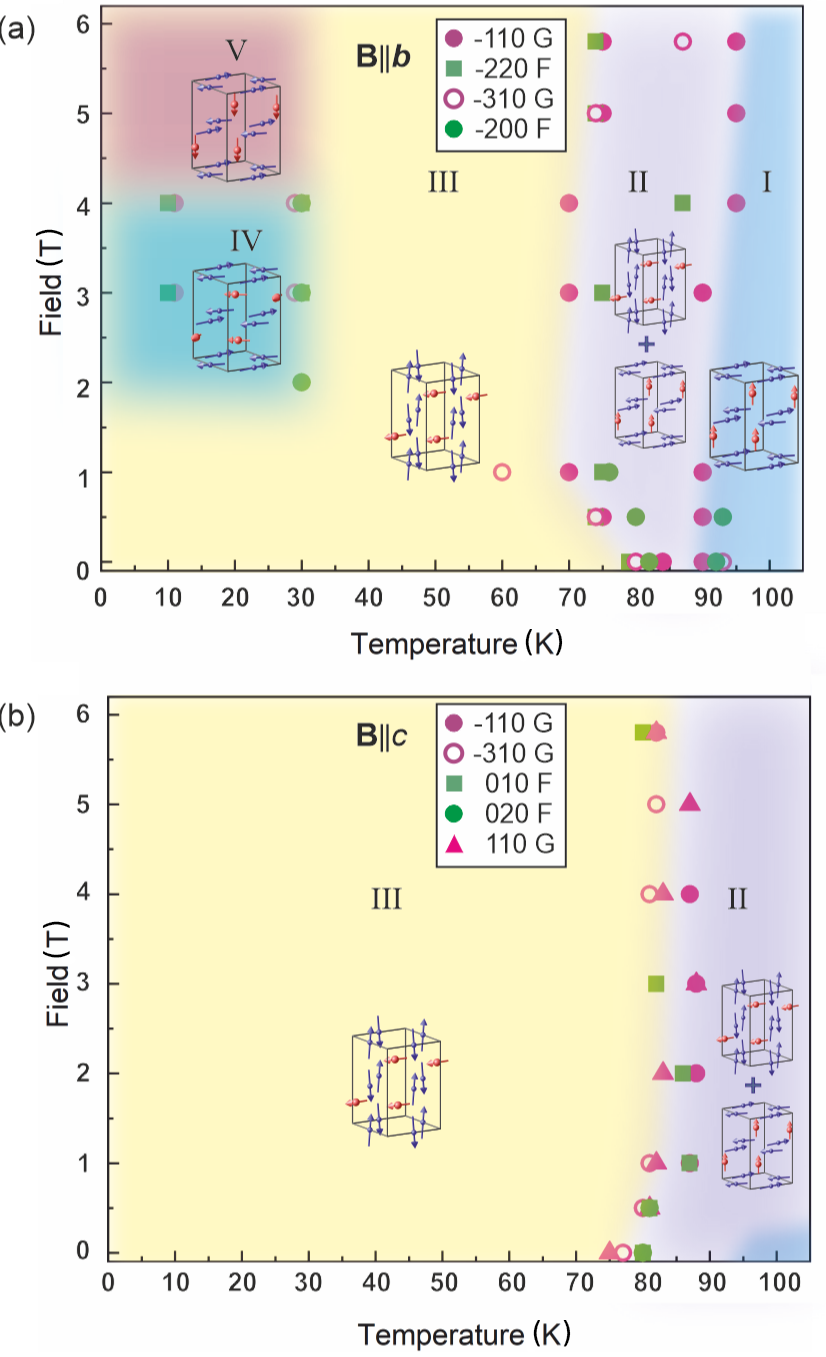}
\caption{\label{fig:fig3} Schematic of the magnetic phase diagram for $\rm{TmFeO_3}$ a) external magnetic field along the \textit{b}-axis b) along \textit{c}-axis. Squares, circles and triangles are the inflection points of the temperature dependence function of the intensity for different reflections. Roman numbers designate regions corresponding by different magnetic representation: I -- $\Gamma 4$; II -- $\Gamma 24$; III -- $\Gamma 2$; IV -- $\Gamma^{\rm{Fe}}_4 + \Gamma^{\rm{Tm}}_2$; V -- $\Gamma^{\rm{Fe}}_4 - \Gamma^{\rm{Tm}}_4$. Magnetic configurations, corresponding each region, are shown schematically without magnetic moments scaling.}
\end{figure}

On the base of our observations, the magnetic phase diagram for $\rm{TmFeO_3}$ in an external fields directed along the \textit{b}-axis and the \textit{c}-axis is built and presented in Figure~\ref{fig:fig3}. Our data did not allow us to define phase boundaries with sufficient accuracy, so the boundaries between phases are depicted rather vaguely. The points on the diagram are defined as the inflection points of the temperature dependence of the intensity. The temperature region 70 -- 90~K in Figure~\ref{fig:fig3}a corresponds to the mixed phase $\Gamma 24$, which arises at the spin reorientation transition. Magnetic structure and spin reorientation in $\rm{TmFeO_3}$ without magnetic field was examined in~\cite{Leake1968}. As it is generally accepted, spin-reorientation in $\rm{RFeO_3}$ is closely connected with Fe-R interaction. At lower temperatures the $\rm{R^{3+}}$ ions become polarized, and at some moment the anisotropy of rare-earth subsystem overcomes the intrinsic anisotropy of the Fe system which prefers the $\Gamma 4$ ($A_x, F_y, G_z$) configuration. Thus new interaction between 4\textit{f} and 3\textit{d} magnetic sublattices forces Fe subsystem to reorient to $\Gamma 2$ configuration. It should be noted that, within the framework of recent theoretical models, the spin reorientation transitions could be connected with more complex mechanism that results from the competition between the 4\textit{f}-3\textit{d}, second- and fourth-order interactions between the spin anisotropy of the 3\textit{d} sublattice and the crystal field on rare-earth 4\textit{f}-ions~\cite{Moskvin2022}.

In the case when the magnetic field is directed along the \textit{c}-axis, one can see that the temperature of the transition from the mixed phase $\Gamma 24$ to phase $\Gamma 2$ increases (Figure~\ref{fig:fig3}b).  When the field applied along \textit{c}-axis, which is an antiferromagnetic axis in $\Gamma 4$ phase, it forces the Fe spins to be perpendicular to the field at some critical value $H_{flop}$. This latter depends on anisotropy and exchange fields, and, as one can suppose, it has the value not higher than 1~T. Thus, field applied along \textit{c}-axis favors $\Gamma 2$ mode to be stabilized with lower exchange energy $J^{\rm{Fe-Tm}}$, that leads to the increase of $T_{SR}$. So, field in \textit{c}-direction increases the temperature of spin-reorientation to $\Gamma 2$ configuration, in opposite to the case of field along \textit{b}-axis. As one can see from Figure~\ref{fig:fig3}a the temperature range of the mixed phase $\Gamma 24$ expands to a lower temperatures. In that case, when the field directed along \textit{b}-axis, that is along weak ferromagnetic component in $\Gamma 4$, it maintains $\Gamma 4$ configuration, thus spreading region of mixed phase $\Gamma 24$ to the lower temperatures. Similar shifts of spin reorientation temperature depending on direction of magnetic field was observed in $\rm{DyFeO_3}$~\cite{Zhao2014, Eremenko1987}. In $\rm{TmFeO_3}$ above $T_{SR}$ rotation of weak ferromagnetic moment in \textit{bc} plane when magnetic field is applied along \textit{c}-axis was confirmed using terahertz time-domain spectroscopy~\cite{Guo2020}.

Regarding temperature region 2 -- 30~K and field range 2 -- 4~T, one can suppose, that in this case external magnetic field favors for the decoupling between $\rm{Tm^{3+}}$ and $\rm{Fe^{3+}}$ ions and Fe subsystem recover high-temperature configuration $\Gamma 4$, while Tm subsystem remains in $\Gamma 2$. At higher fields like 5 -- 5.8~T Tm subsystem also goes to $\Gamma 4$ configuration but with ferromagnetic component directed opposite to $\rm{Fe^{3+}}$ one. The change of peaks intensity at magnetic field applied along \textit{b}-axis and the consequent refinement of the magnetic structure confirm supposed phase transitions. Authors of~\cite{Wang2023} observed a magnetic-field-induced spin reorientation $\Gamma 2$ $\rightarrow$ $\Gamma 4$ at 1.6~K in a \textit{b}-cut $\rm{TmFeO_3}$ single crystal in the static magnetic field range 2.2 -- 3.6~T applied along \textit{b}-axis using terahertz spectroscopy. We have also observed change of intensity of peaks at magnetic field applied along \textit{b}-axis in the close field range that confirms phase transition. Situation with reverse reorientation transition without external magnetic field was observed in $\rm{TbFeO_3}$, a unique compound in which two types of spin reorientation from $\Gamma 4$ to $\Gamma 2$ at 8.5~K and then $\Gamma 2$ to $\Gamma 4$ at 3~K were reported~\cite{Belov1979}.

\begin{figure} 
\includegraphics[width=0.45\textwidth]{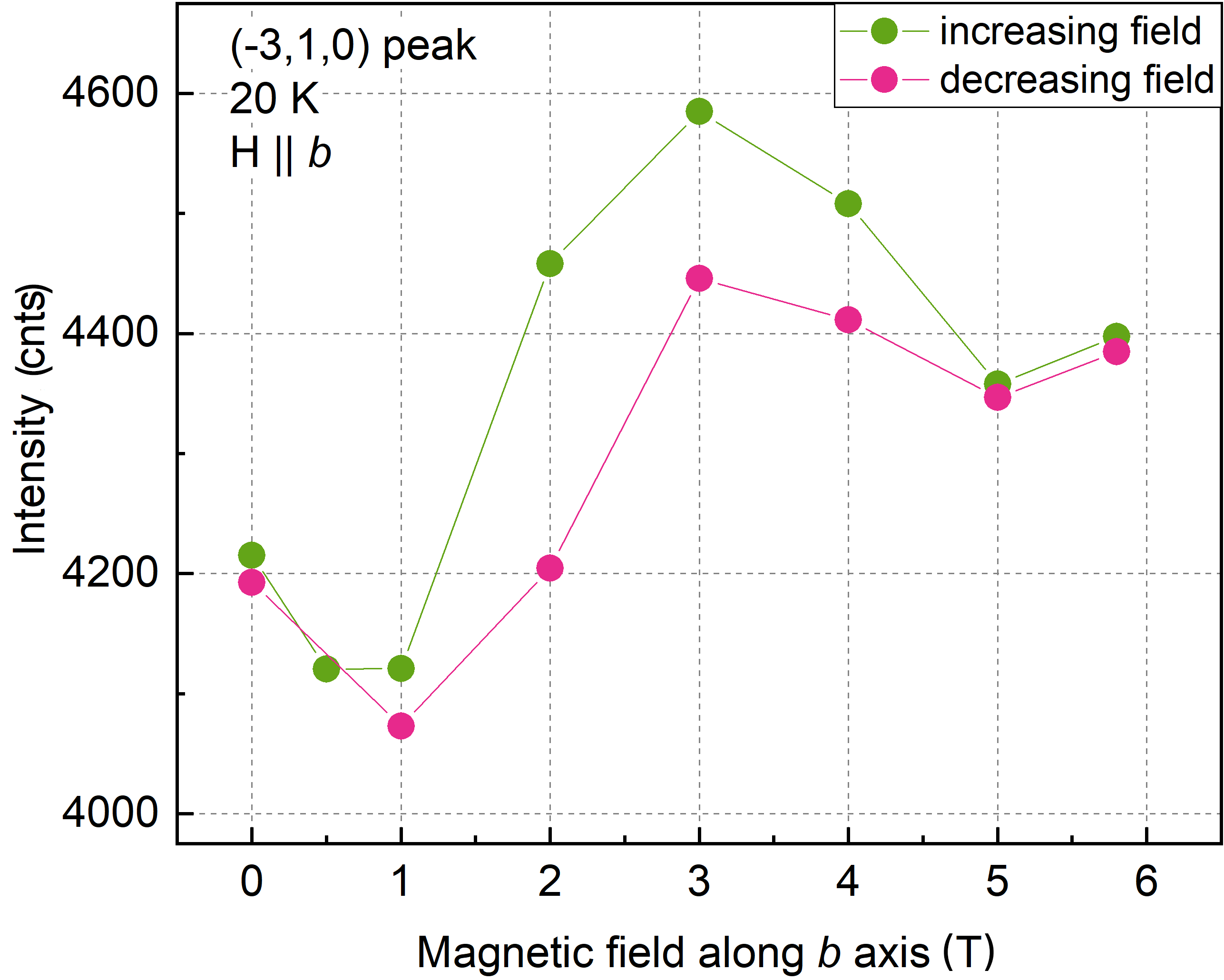}
\caption{\label{fig:fig4} Hysteresis of G-type peak ($-3,1,0$) intensity in magnetic fields along the \textit{b}-axis at 20~K. Error bars are of the size of the symbols.}
\end{figure}

In addition, hysteresis of the peak’s intensity for the field along the \textit{b}-axis was observed in the experiment (Fig.~\ref{fig:fig4}), which was absent with the field along the \textit{c}-axis, i.e. along the giant magnetocaloric axis. The largest difference between the peak intensities measured when the fields were increased and decreased was observed for G-type peaks like $(-3,1,0)$, presented at Figure~\ref{fig:fig4}, as it is provided by the strongest G-type ordered component of the magnetic structure. A smaller hysteresis was observed for A-type and F-type peaks, that connected with the smaller magnetic peaks’ intensities. Apparently, this difference is related with domains transformation when going from magnetic phase IV to phase V and back, which phases emerge when magnetic field applied along the \textit{b}-axis. The domain structure in $\rm{TmFeO_3}$ was confirmed in work~\cite{Tsymbal2010} where temperature magnetization hysteresis was described using model with a few domains, which explain jump/drop of magnetization in the spin reorientation temperatures range.

\section{Conclusions}
X-ray studies demonstrate that the lattice parameters and atomic coordinates change almost linearly in $\rm{TmFeO_3}$ through the spin reorientation transition, keeping the orthorhombic symmetry group \textit{Pnma}. The direction of the electric dipole moments does not change at any magnetic phase. This is additional evidence that electric system of $\rm{TmFeO_3}$ does not have influence on the spin-reorientation transition in this compound. The main reason for the spin reorientation transition connected with superexchange interactions 4\textit{f}-3\textit{d}, or with more  complicate scheme of magnetic interactions~\cite{Moskvin2022}. Application of external magnetic field influences significantly on the spin-reorientation processes, moving the temperatures of the transitions up or low, depending on the direction of the external field. Thus, the application of the field along the c axis creates a situation favorable for the existence of the $\Gamma 2$ configuration, which leads to an increase in temperature of transition from $\Gamma 4$ to $\Gamma 2$ phase. The field along \textit{b}-axis of $\rm{TmFeO_3}$ favors for $\Gamma 4$ configuration. This results in a decrease in the transition temperature from the $\Gamma 4$ phase to the $\Gamma 2$ phase, an increase in the temperature range in which the mixed phase is observed; and also allows for the formation of new phases. At low temperatures below 20~K and fields between 2 -- 4~T a new magnetic configuration was observed, which consists of representation $\Gamma 4$ for $\rm{Fe^{3+}}$ ions and $\Gamma 2$ for $\rm{Tm^{3+}}$. At higher fields at these low temperatures another magnetic configuration emerges with representation $\Gamma 4$ for both $\rm{Fe^{3+}}$ subsystem and $\Gamma 4$ for $\rm{Tm^{3+}}$, but with opposite signs for $F_z$-components.

The temperature region 2 -- 30~K and field range 2 -- 4~T of formation new phase are in correlation with the data on the measurement of the magnetocaloric effect, where the peak corresponding to the maximum of magnetic entropy change $\Delta S \approx 12$~J/kg$\cdot$K lies in the range of $T = 15 - 25$~K~\cite{Ke2016}.

\begin{acknowledgments}
 We acknowledge Andrew Wildes (ILL) for help with crystal orientation using OrientExpress. This work was supported by the DFG grant \#SA 3688/1-1. The synthesis of single crystal $\rm{TmFeO_3}$ was carried out within the State assignment of the Ministry of Science and Higher Education of the Russian Federation for the Kirensky Institute of Physics, Krasnoyarsk Scientific Center, Siberian Branch of the Russian Academy of Sciences.
\end{acknowledgments}


\bibliography{main_ref}

\providecommand{\noopsort}[1]{}\providecommand{\singleletter}[1]{#1}%
\begin{thebibliography}{45}%
\makeatletter
\providecommand \@ifxundefined [1]{%
 \@ifx{#1\undefined}
}%
\providecommand \@ifnum [1]{%
 \ifnum #1\expandafter \@firstoftwo
 \else \expandafter \@secondoftwo
 \fi
}%
\providecommand \@ifx [1]{%
 \ifx #1\expandafter \@firstoftwo
 \else \expandafter \@secondoftwo
 \fi
}%
\providecommand \natexlab [1]{#1}%
\providecommand \enquote  [1]{``#1''}%
\providecommand \bibnamefont  [1]{#1}%
\providecommand \bibfnamefont [1]{#1}%
\providecommand \citenamefont [1]{#1}%
\providecommand \href@noop [0]{\@secondoftwo}%
\providecommand \href [0]{\begingroup \@sanitize@url \@href}%
\providecommand \@href[1]{\@@startlink{#1}\@@href}%
\providecommand \@@href[1]{\endgroup#1\@@endlink}%
\providecommand \@sanitize@url [0]{\catcode `\\12\catcode `\$12\catcode
  `\&12\catcode `\#12\catcode `\^12\catcode `\_12\catcode `\%12\relax}%
\providecommand \@@startlink[1]{}%
\providecommand \@@endlink[0]{}%
\providecommand \url  [0]{\begingroup\@sanitize@url \@url }%
\providecommand \@url [1]{\endgroup\@href {#1}{\urlprefix }}%
\providecommand \urlprefix  [0]{URL }%
\providecommand \Eprint [0]{\href }%
\providecommand \doibase [0]{https://doi.org/}%
\providecommand \selectlanguage [0]{\@gobble}%
\providecommand \bibinfo  [0]{\@secondoftwo}%
\providecommand \bibfield  [0]{\@secondoftwo}%
\providecommand \translation [1]{[#1]}%
\providecommand \BibitemOpen [0]{}%
\providecommand \bibitemStop [0]{}%
\providecommand \bibitemNoStop [0]{.\EOS\space}%
\providecommand \EOS [0]{\spacefactor3000\relax}%
\providecommand \BibitemShut  [1]{\csname bibitem#1\endcsname}%
\let\auto@bib@innerbib\@empty
\bibitem [{\citenamefont {White}(1969)}]{White1969}%
  \BibitemOpen
  \bibfield  {author} {\bibinfo {author} {\bibfnamefont {R.}~\bibnamefont
  {White}},\ }\bibfield  {title} {\bibinfo {title} {Review of recent work on
  the magnetic and spectroscopic properties of the rare-earth orthoferrites},\
  }\href {https://doi.org/10.1063/1.1657530} {\bibfield  {journal} {\bibinfo
  {journal} {Journal of Applied Physics}\ }\textbf {\bibinfo {volume} {40}},\
  \bibinfo {pages} {1061} (\bibinfo {year} {1969})}\BibitemShut {NoStop}%
\bibitem [{\citenamefont {Dzyaloshinsky}(1958)}]{Dzyaloshinsky1958}%
  \BibitemOpen
  \bibfield  {author} {\bibinfo {author} {\bibfnamefont {I.}~\bibnamefont
  {Dzyaloshinsky}},\ }\bibfield  {title} {\bibinfo {title} {A thermodynamic
  theory of “weak” ferromagnetism of antiferromagnetics},\ }\href
  {https://doi.org/10.1016/0022-3697(58)90076-3} {\bibfield  {journal}
  {\bibinfo  {journal} {Journal of Physics and Chemistry of Solids}\ }\textbf
  {\bibinfo {volume} {4}},\ \bibinfo {pages} {241} (\bibinfo {year}
  {1958})}\BibitemShut {NoStop}%
\bibitem [{\citenamefont {Moriya}(1960)}]{Moriya1960}%
  \BibitemOpen
  \bibfield  {author} {\bibinfo {author} {\bibfnamefont {T.}~\bibnamefont
  {Moriya}},\ }\bibfield  {title} {\bibinfo {title} {Anisotropic superexchange
  interaction and weak ferromagnetism},\ }\href
  {https://doi.org/10.1103/physrev.120.91} {\bibfield  {journal} {\bibinfo
  {journal} {Physical Review}\ }\textbf {\bibinfo {volume} {120}},\ \bibinfo
  {pages} {91} (\bibinfo {year} {1960})}\BibitemShut {NoStop}%
\bibitem [{\citenamefont {Artyukhin}\ \emph {et~al.}(2012)\citenamefont
  {Artyukhin}, \citenamefont {Mostovoy}, \citenamefont {Jensen}, \citenamefont
  {Le}, \citenamefont {Prokes}, \citenamefont {De~Paula}, \citenamefont
  {Bordallo}, \citenamefont {Maljuk}, \citenamefont {Landsgesell},
  \citenamefont {Ryll} \emph {et~al.}}]{Artyukhin2012}%
  \BibitemOpen
  \bibfield  {author} {\bibinfo {author} {\bibfnamefont {S.}~\bibnamefont
  {Artyukhin}}, \bibinfo {author} {\bibfnamefont {M.}~\bibnamefont {Mostovoy}},
  \bibinfo {author} {\bibfnamefont {N.~P.}\ \bibnamefont {Jensen}}, \bibinfo
  {author} {\bibfnamefont {D.}~\bibnamefont {Le}}, \bibinfo {author}
  {\bibfnamefont {K.}~\bibnamefont {Prokes}}, \bibinfo {author} {\bibfnamefont
  {V.~G.}\ \bibnamefont {De~Paula}}, \bibinfo {author} {\bibfnamefont {H.~N.}\
  \bibnamefont {Bordallo}}, \bibinfo {author} {\bibfnamefont {A.}~\bibnamefont
  {Maljuk}}, \bibinfo {author} {\bibfnamefont {S.}~\bibnamefont {Landsgesell}},
  \bibinfo {author} {\bibfnamefont {H.}~\bibnamefont {Ryll}}, \emph {et~al.},\
  }\bibfield  {title} {\bibinfo {title} {Solitonic lattice and yukawa forces in
  the rare-earth orthoferrite $\rm{TbFeO_3}$},\ }\href
  {https://doi.org/10.1038/nmat3358} {\bibfield  {journal} {\bibinfo  {journal}
  {Nature Materials}\ }\textbf {\bibinfo {volume} {11}},\ \bibinfo {pages}
  {694} (\bibinfo {year} {2012})}\BibitemShut {NoStop}%
\bibitem [{\citenamefont {Nikolov}\ \emph {et~al.}(1994)\citenamefont
  {Nikolov}, \citenamefont {Hall}, \citenamefont {Barilo},\ and\ \citenamefont
  {Guretskii}}]{Nikolov1994}%
  \BibitemOpen
  \bibfield  {author} {\bibinfo {author} {\bibfnamefont {O.}~\bibnamefont
  {Nikolov}}, \bibinfo {author} {\bibfnamefont {I.}~\bibnamefont {Hall}},
  \bibinfo {author} {\bibfnamefont {S.}~\bibnamefont {Barilo}},\ and\ \bibinfo
  {author} {\bibfnamefont {S.}~\bibnamefont {Guretskii}},\ }\bibfield  {title}
  {\bibinfo {title} {A mossbauer study of temperature-driven spin-reorientation
  transitions in $\rm{TbFeO_3}$},\ }\href
  {https://doi.org/10.1088/0953-8984/6/20/019} {\bibfield  {journal} {\bibinfo
  {journal} {Journal of Physics: Condensed Matter}\ }\textbf {\bibinfo {volume}
  {6}},\ \bibinfo {pages} {3793} (\bibinfo {year} {1994})}\BibitemShut
  {NoStop}%
\bibitem [{\citenamefont {Tokunaga}\ \emph {et~al.}(2008)\citenamefont
  {Tokunaga}, \citenamefont {Iguchi}, \citenamefont {Arima},\ and\
  \citenamefont {Tokura}}]{Tokunaga2008}%
  \BibitemOpen
  \bibfield  {author} {\bibinfo {author} {\bibfnamefont {Y.}~\bibnamefont
  {Tokunaga}}, \bibinfo {author} {\bibfnamefont {S.}~\bibnamefont {Iguchi}},
  \bibinfo {author} {\bibfnamefont {T.-H.}\ \bibnamefont {Arima}},\ and\
  \bibinfo {author} {\bibfnamefont {Y.}~\bibnamefont {Tokura}},\ }\bibfield
  {title} {\bibinfo {title} {Magnetic-field-induced ferroelectric state in
  $\rm{DyFeO_3}$},\ }\href {https://doi.org/10.1103/physrevlett.101.097205}
  {\bibfield  {journal} {\bibinfo  {journal} {Physical Review Letters}\
  }\textbf {\bibinfo {volume} {101}},\ \bibinfo {pages} {097205} (\bibinfo
  {year} {2008})}\BibitemShut {NoStop}%
\bibitem [{\citenamefont {Tokunaga}\ \emph {et~al.}(2009)\citenamefont
  {Tokunaga}, \citenamefont {Furukawa}, \citenamefont {Sakai}, \citenamefont
  {Taguchi}, \citenamefont {Arima},\ and\ \citenamefont
  {Tokura}}]{Tokunaga2009}%
  \BibitemOpen
  \bibfield  {author} {\bibinfo {author} {\bibfnamefont {Y.}~\bibnamefont
  {Tokunaga}}, \bibinfo {author} {\bibfnamefont {N.}~\bibnamefont {Furukawa}},
  \bibinfo {author} {\bibfnamefont {H.}~\bibnamefont {Sakai}}, \bibinfo
  {author} {\bibfnamefont {Y.}~\bibnamefont {Taguchi}}, \bibinfo {author}
  {\bibfnamefont {T.-h.}\ \bibnamefont {Arima}},\ and\ \bibinfo {author}
  {\bibfnamefont {Y.}~\bibnamefont {Tokura}},\ }\bibfield  {title} {\bibinfo
  {title} {Composite domain walls in a multiferroic perovskite ferrite},\
  }\href {https://doi.org/10.1038/nmat2469} {\bibfield  {journal} {\bibinfo
  {journal} {Nature Materials}\ }\textbf {\bibinfo {volume} {8}},\ \bibinfo
  {pages} {558} (\bibinfo {year} {2009})}\BibitemShut {NoStop}%
\bibitem [{\citenamefont {Song}\ \emph {et~al.}(2014)\citenamefont {Song},
  \citenamefont {Zhou}, \citenamefont {Fang}, \citenamefont {Yang},
  \citenamefont {Wang}, \citenamefont {Wang},\ and\ \citenamefont
  {Du}}]{Song2014}%
  \BibitemOpen
  \bibfield  {author} {\bibinfo {author} {\bibfnamefont {Y.-Q.}\ \bibnamefont
  {Song}}, \bibinfo {author} {\bibfnamefont {W.-P.}\ \bibnamefont {Zhou}},
  \bibinfo {author} {\bibfnamefont {Y.}~\bibnamefont {Fang}}, \bibinfo {author}
  {\bibfnamefont {Y.-T.}\ \bibnamefont {Yang}}, \bibinfo {author}
  {\bibfnamefont {L.-Y.}\ \bibnamefont {Wang}}, \bibinfo {author}
  {\bibfnamefont {D.-H.}\ \bibnamefont {Wang}},\ and\ \bibinfo {author}
  {\bibfnamefont {Y.-W.}\ \bibnamefont {Du}},\ }\bibfield  {title} {\bibinfo
  {title} {Multiferroic properties in terbium orthoferrite},\ }\href
  {https://doi.org/10.1088/1674-1056/23/7/077505} {\bibfield  {journal}
  {\bibinfo  {journal} {Chinese Physics B}\ }\textbf {\bibinfo {volume} {23}},\
  \bibinfo {pages} {077505} (\bibinfo {year} {2014})}\BibitemShut {NoStop}%
\bibitem [{\citenamefont {Lee}\ \emph {et~al.}(2011)\citenamefont {Lee},
  \citenamefont {Jeong}, \citenamefont {Park}, \citenamefont {Oak},
  \citenamefont {Jang}, \citenamefont {Son},\ and\ \citenamefont
  {Scott}}]{Lee2011}%
  \BibitemOpen
  \bibfield  {author} {\bibinfo {author} {\bibfnamefont {J.-H.}\ \bibnamefont
  {Lee}}, \bibinfo {author} {\bibfnamefont {Y.~K.}\ \bibnamefont {Jeong}},
  \bibinfo {author} {\bibfnamefont {J.~H.}\ \bibnamefont {Park}}, \bibinfo
  {author} {\bibfnamefont {M.-A.}\ \bibnamefont {Oak}}, \bibinfo {author}
  {\bibfnamefont {H.~M.}\ \bibnamefont {Jang}}, \bibinfo {author}
  {\bibfnamefont {J.~Y.}\ \bibnamefont {Son}},\ and\ \bibinfo {author}
  {\bibfnamefont {J.~F.}\ \bibnamefont {Scott}},\ }\bibfield  {title} {\bibinfo
  {title} {Spin-canting-induced improper ferroelectricity and spontaneous
  magnetization reversal in $\rm{SmFeO_3}$},\ }\href
  {https://doi.org/10.1103/physrevlett.107.117201} {\bibfield  {journal}
  {\bibinfo  {journal} {Physical Review Letters}\ }\textbf {\bibinfo {volume}
  {107}},\ \bibinfo {pages} {117201} (\bibinfo {year} {2011})}\BibitemShut
  {NoStop}%
\bibitem [{\citenamefont {Shang}\ \emph {et~al.}(2013)\citenamefont {Shang},
  \citenamefont {Zhang}, \citenamefont {Zhang}, \citenamefont {Yuan},
  \citenamefont {Ge}, \citenamefont {Yuan},\ and\ \citenamefont
  {Feng}}]{Shang2013}%
  \BibitemOpen
  \bibfield  {author} {\bibinfo {author} {\bibfnamefont {M.}~\bibnamefont
  {Shang}}, \bibinfo {author} {\bibfnamefont {C.}~\bibnamefont {Zhang}},
  \bibinfo {author} {\bibfnamefont {T.}~\bibnamefont {Zhang}}, \bibinfo
  {author} {\bibfnamefont {L.}~\bibnamefont {Yuan}}, \bibinfo {author}
  {\bibfnamefont {L.}~\bibnamefont {Ge}}, \bibinfo {author} {\bibfnamefont
  {H.}~\bibnamefont {Yuan}},\ and\ \bibinfo {author} {\bibfnamefont
  {S.}~\bibnamefont {Feng}},\ }\bibfield  {title} {\bibinfo {title} {The
  multiferroic perovskite $\rm{YFeO_3}$},\ }\bibfield  {journal} {\bibinfo
  {journal} {Applied Physics Letters}\ }\textbf {\bibinfo {volume} {102}},\
  \href {https://doi.org/10.1063/1.4791697} {10.1063/1.4791697} (\bibinfo
  {year} {2013})\BibitemShut {NoStop}%
\bibitem [{\citenamefont {Kimel}\ \emph {et~al.}(2004)\citenamefont {Kimel},
  \citenamefont {Kirilyuk}, \citenamefont {Tsvetkov}, \citenamefont {Pisarev},\
  and\ \citenamefont {Rasing}}]{Kimel2004}%
  \BibitemOpen
  \bibfield  {author} {\bibinfo {author} {\bibfnamefont {A.}~\bibnamefont
  {Kimel}}, \bibinfo {author} {\bibfnamefont {A.}~\bibnamefont {Kirilyuk}},
  \bibinfo {author} {\bibfnamefont {A.}~\bibnamefont {Tsvetkov}}, \bibinfo
  {author} {\bibfnamefont {R.}~\bibnamefont {Pisarev}},\ and\ \bibinfo {author}
  {\bibfnamefont {T.}~\bibnamefont {Rasing}},\ }\bibfield  {title} {\bibinfo
  {title} {Laser-induced ultrafast spin reorientation in the antiferromagnet
  $\rm{TmFeO_3}$},\ }\href {https://doi.org/10.1038/nature02659} {\bibfield
  {journal} {\bibinfo  {journal} {Nature}\ }\textbf {\bibinfo {volume} {429}},\
  \bibinfo {pages} {850} (\bibinfo {year} {2004})}\BibitemShut {NoStop}%
\bibitem [{\citenamefont {Bombik}\ \emph {et~al.}(2003)\citenamefont {Bombik},
  \citenamefont {Le{\'s}niewska}, \citenamefont {Mayer},\ and\ \citenamefont
  {Pacyna}}]{Bombik2003}%
  \BibitemOpen
  \bibfield  {author} {\bibinfo {author} {\bibfnamefont {A.}~\bibnamefont
  {Bombik}}, \bibinfo {author} {\bibfnamefont {B.}~\bibnamefont
  {Le{\'s}niewska}}, \bibinfo {author} {\bibfnamefont {J.}~\bibnamefont
  {Mayer}},\ and\ \bibinfo {author} {\bibfnamefont {A.~W.}\ \bibnamefont
  {Pacyna}},\ }\bibfield  {title} {\bibinfo {title} {Crystal structure of solid
  solutions $\rm{REFe_{1-x}(Al~\text{or}~ Ga)_xO_3}$ $\rm{(RE = Tb,~Er,~Tm)}$
  and the correlation between superexchange interaction $\rm{Fe^{+3} - O^{-2} -
  Fe^{+3}}$ linkage angles and \uppercase{N}{\'e}el temperature},\ }\href
  {https://doi.org/10.1016/s0304-8853(02)01172-1} {\bibfield  {journal}
  {\bibinfo  {journal} {Journal of Magnetism and Magnetic Materials}\ }\textbf
  {\bibinfo {volume} {257}},\ \bibinfo {pages} {206} (\bibinfo {year}
  {2003})}\BibitemShut {NoStop}%
\bibitem [{\citenamefont {Bertaut}(1963)}]{Bertaut1963}%
  \BibitemOpen
  \bibfield  {author} {\bibinfo {author} {\bibfnamefont {E.~F.}\ \bibnamefont
  {Bertaut}},\ }\href@noop {} {\emph {\bibinfo {title} {Magnetism}}}\ (\bibinfo
   {publisher} {Academic},\ \bibinfo {address} {New York},\ \bibinfo {year}
  {1963})\BibitemShut {NoStop}%
\bibitem [{\citenamefont {Leake}\ \emph {et~al.}(1968)\citenamefont {Leake},
  \citenamefont {Shirane},\ and\ \citenamefont {Remeika}}]{Leake1968}%
  \BibitemOpen
  \bibfield  {author} {\bibinfo {author} {\bibfnamefont {J.}~\bibnamefont
  {Leake}}, \bibinfo {author} {\bibfnamefont {G.}~\bibnamefont {Shirane}},\
  and\ \bibinfo {author} {\bibfnamefont {J.}~\bibnamefont {Remeika}},\
  }\bibfield  {title} {\bibinfo {title} {The magnetic structure of thulium
  orthoferrite, $\rm{TmFeO_3}$},\ }\href@noop {} {\bibfield  {journal}
  {\bibinfo  {journal} {Solid State Communications}\ }\textbf {\bibinfo
  {volume} {6}},\ \bibinfo {pages} {15} (\bibinfo {year} {1968})}\BibitemShut
  {NoStop}%
\bibitem [{\citenamefont {Yamaguchi}\ and\ \citenamefont
  {Tsushima}(1973)}]{Tsuyoshi1973}%
  \BibitemOpen
  \bibfield  {author} {\bibinfo {author} {\bibfnamefont {T.}~\bibnamefont
  {Yamaguchi}}\ and\ \bibinfo {author} {\bibfnamefont {K.}~\bibnamefont
  {Tsushima}},\ }\bibfield  {title} {\bibinfo {title} {Magnetic symmetry of
  rare-earth orthochromites and orthoferrites},\ }\href
  {https://doi.org/10.1103/physrevb.8.5187} {\bibfield  {journal} {\bibinfo
  {journal} {Physical Review B}\ }\textbf {\bibinfo {volume} {8}},\ \bibinfo
  {pages} {5187} (\bibinfo {year} {1973})}\BibitemShut {NoStop}%
\bibitem [{\citenamefont {Ke}\ \emph {et~al.}(2016)\citenamefont {Ke},
  \citenamefont {Zhang}, \citenamefont {Ma},\ and\ \citenamefont
  {Cheng}}]{Ke2016}%
  \BibitemOpen
  \bibfield  {author} {\bibinfo {author} {\bibfnamefont {Y.-J.}\ \bibnamefont
  {Ke}}, \bibinfo {author} {\bibfnamefont {X.-Q.}\ \bibnamefont {Zhang}},
  \bibinfo {author} {\bibfnamefont {Y.}~\bibnamefont {Ma}},\ and\ \bibinfo
  {author} {\bibfnamefont {Z.-H.}\ \bibnamefont {Cheng}},\ }\bibfield  {title}
  {\bibinfo {title} {Anisotropic magnetic entropy change in $\rm{RFeO_3}$
  single crystals $\rm{(R=Tb,~Tm,~\text{or}~Y)}$},\ }\href
  {https://doi.org/10.1038/srep19775} {\bibfield  {journal} {\bibinfo
  {journal} {Scientific Reports}\ }\textbf {\bibinfo {volume} {6}},\ \bibinfo
  {pages} {19775} (\bibinfo {year} {2016})}\BibitemShut {NoStop}%
\bibitem [{\citenamefont {Skorobogatov}\ \emph {et~al.}(2020)\citenamefont
  {Skorobogatov}, \citenamefont {Nikitin}, \citenamefont {Shaykhutdinov},
  \citenamefont {Balaev}, \citenamefont {Terentjev}, \citenamefont {Ehlers},
  \citenamefont {Sala}, \citenamefont {Pomjakushina}, \citenamefont {Conder},\
  and\ \citenamefont {Podlesnyak}}]{Skorobogatov2020}%
  \BibitemOpen
  \bibfield  {author} {\bibinfo {author} {\bibfnamefont {S.}~\bibnamefont
  {Skorobogatov}}, \bibinfo {author} {\bibfnamefont {S.~E.}\ \bibnamefont
  {Nikitin}}, \bibinfo {author} {\bibfnamefont {K.}~\bibnamefont
  {Shaykhutdinov}}, \bibinfo {author} {\bibfnamefont {A.}~\bibnamefont
  {Balaev}}, \bibinfo {author} {\bibfnamefont {K.~Y.}\ \bibnamefont
  {Terentjev}}, \bibinfo {author} {\bibfnamefont {G.}~\bibnamefont {Ehlers}},
  \bibinfo {author} {\bibfnamefont {G.}~\bibnamefont {Sala}}, \bibinfo {author}
  {\bibfnamefont {E.}~\bibnamefont {Pomjakushina}}, \bibinfo {author}
  {\bibfnamefont {K.}~\bibnamefont {Conder}},\ and\ \bibinfo {author}
  {\bibfnamefont {A.}~\bibnamefont {Podlesnyak}},\ }\bibfield  {title}
  {\bibinfo {title} {Low-temperature spin dynamics in the $\rm{TmFeO_3}$
  orthoferrite with a non-kramers ion},\ }\href
  {https://doi.org/10.1103/physrevb.101.014432} {\bibfield  {journal} {\bibinfo
   {journal} {Physical Review B}\ }\textbf {\bibinfo {volume} {101}},\ \bibinfo
  {pages} {014432} (\bibinfo {year} {2020})}\BibitemShut {NoStop}%
\bibitem [{\citenamefont {Ovsianikov}\ \emph {et~al.}(2022)\citenamefont
  {Ovsianikov}, \citenamefont {Usmanov}, \citenamefont {Zobkalo}, \citenamefont
  {Schmidt}, \citenamefont {Maity}, \citenamefont {Hutanu}, \citenamefont
  {Ressouche}, \citenamefont {Shaykhutdinov}, \citenamefont {Terentjev},
  \citenamefont {Semenov} \emph {et~al.}}]{Ovsianikov2022}%
  \BibitemOpen
  \bibfield  {author} {\bibinfo {author} {\bibfnamefont {A.}~\bibnamefont
  {Ovsianikov}}, \bibinfo {author} {\bibfnamefont {O.}~\bibnamefont {Usmanov}},
  \bibinfo {author} {\bibfnamefont {I.}~\bibnamefont {Zobkalo}}, \bibinfo
  {author} {\bibfnamefont {W.}~\bibnamefont {Schmidt}}, \bibinfo {author}
  {\bibfnamefont {A.}~\bibnamefont {Maity}}, \bibinfo {author} {\bibfnamefont
  {V.}~\bibnamefont {Hutanu}}, \bibinfo {author} {\bibfnamefont
  {E.}~\bibnamefont {Ressouche}}, \bibinfo {author} {\bibfnamefont
  {K.}~\bibnamefont {Shaykhutdinov}}, \bibinfo {author} {\bibfnamefont {K.~Y.}\
  \bibnamefont {Terentjev}}, \bibinfo {author} {\bibfnamefont {S.}~\bibnamefont
  {Semenov}}, \emph {et~al.},\ }\bibfield  {title} {\bibinfo {title} {Inelastic
  neutron studies and diffraction in magnetic fields of $\rm{TbFeO_3}$ and
  $\rm{YbFeO_3}$},\ }\href {https://doi.org/10.1016/j.jmmm.2022.170025}
  {\bibfield  {journal} {\bibinfo  {journal} {Journal of Magnetism and Magnetic
  Materials}\ }\textbf {\bibinfo {volume} {563}},\ \bibinfo {pages} {170025}
  (\bibinfo {year} {2022})}\BibitemShut {NoStop}%
\bibitem [{\citenamefont {Fabrykiewicz}\ \emph {et~al.}(2021)\citenamefont
  {Fabrykiewicz}, \citenamefont {Przenios{\l}o},\ and\ \citenamefont
  {Sosnowska}}]{Fabrykiewicz2021}%
  \BibitemOpen
  \bibfield  {author} {\bibinfo {author} {\bibfnamefont {P.}~\bibnamefont
  {Fabrykiewicz}}, \bibinfo {author} {\bibfnamefont {R.}~\bibnamefont
  {Przenios{\l}o}},\ and\ \bibinfo {author} {\bibfnamefont {I.}~\bibnamefont
  {Sosnowska}},\ }\bibfield  {title} {\bibinfo {title} {Magnetic modes
  compatible with the symmetry of crystals},\ }\href@noop {} {\bibfield
  {journal} {\bibinfo  {journal} {Acta Crystallographica Section A: Foundations
  and Advances}\ }\textbf {\bibinfo {volume} {77}},\ \bibinfo {pages} {327}
  (\bibinfo {year} {2021})}\BibitemShut {NoStop}%
\bibitem [{\citenamefont {Barilo}\ \emph {et~al.}(1991)\citenamefont {Barilo},
  \citenamefont {Ges}, \citenamefont {Guretskii}, \citenamefont {Zhigunov},
  \citenamefont {Ignatenko}, \citenamefont {Igumentsev}, \citenamefont
  {Lomako},\ and\ \citenamefont {Luginets}}]{Barilo1991}%
  \BibitemOpen
  \bibfield  {author} {\bibinfo {author} {\bibfnamefont {S.}~\bibnamefont
  {Barilo}}, \bibinfo {author} {\bibfnamefont {A.}~\bibnamefont {Ges}},
  \bibinfo {author} {\bibfnamefont {S.}~\bibnamefont {Guretskii}}, \bibinfo
  {author} {\bibfnamefont {D.}~\bibnamefont {Zhigunov}}, \bibinfo {author}
  {\bibfnamefont {A.}~\bibnamefont {Ignatenko}}, \bibinfo {author}
  {\bibfnamefont {A.}~\bibnamefont {Igumentsev}}, \bibinfo {author}
  {\bibfnamefont {I.}~\bibnamefont {Lomako}},\ and\ \bibinfo {author}
  {\bibfnamefont {A.}~\bibnamefont {Luginets}},\ }\bibfield  {title} {\bibinfo
  {title} {Seeded growth of rare-earth orthoferrites from
  $\rm{B_2O_3-BaF_2-BaO}$ solvent ii. growth of high-quality $\rm{RFeO_3}$
  single crystals},\ }\href {https://doi.org/10.1016/0022-0248(91)90379-j}
  {\bibfield  {journal} {\bibinfo  {journal} {Journal of Crystal Growth}\
  }\textbf {\bibinfo {volume} {108}},\ \bibinfo {pages} {314} (\bibinfo {year}
  {1991})}\BibitemShut {NoStop}%
\bibitem [{\citenamefont {Rodriguez-Cavajal}(2001)}]{RodriguezCarvajal}%
  \BibitemOpen
  \bibfield  {author} {\bibinfo {author} {\bibfnamefont {J.}~\bibnamefont
  {Rodriguez-Cavajal}},\ }\bibfield  {title} {\bibinfo {title} {Recent
  developments of the program fullprof},\ }\href@noop {} {\bibfield  {journal}
  {\bibinfo  {journal} {Comm. Powder Diffract. Newsl.}\ }\textbf {\bibinfo
  {volume} {26}},\ \bibinfo {pages} {12} (\bibinfo {year} {2001})}\BibitemShut
  {NoStop}%
\bibitem [{\citenamefont {Ouladdiaf}\ \emph {et~al.}(2006)\citenamefont
  {Ouladdiaf}, \citenamefont {Archer}, \citenamefont {McIntyre}, \citenamefont
  {Hewat}, \citenamefont {Brau},\ and\ \citenamefont {York}}]{Ouladdiaf2006}%
  \BibitemOpen
  \bibfield  {author} {\bibinfo {author} {\bibfnamefont {B.}~\bibnamefont
  {Ouladdiaf}}, \bibinfo {author} {\bibfnamefont {J.}~\bibnamefont {Archer}},
  \bibinfo {author} {\bibfnamefont {G.}~\bibnamefont {McIntyre}}, \bibinfo
  {author} {\bibfnamefont {A.}~\bibnamefont {Hewat}}, \bibinfo {author}
  {\bibfnamefont {D.}~\bibnamefont {Brau}},\ and\ \bibinfo {author}
  {\bibfnamefont {S.}~\bibnamefont {York}},\ }\bibfield  {title} {\bibinfo
  {title} {Orientexpress: A new system for laue neutron diffraction},\ }\href
  {https://doi.org/10.1016/j.physb.2006.05.337} {\bibfield  {journal} {\bibinfo
   {journal} {Physica B: Condensed Matter}\ }\textbf {\bibinfo {volume}
  {385}},\ \bibinfo {pages} {1052} (\bibinfo {year} {2006})}\BibitemShut
  {NoStop}%
\bibitem [{\citenamefont {{Institut Laue-Langevin}}(2023)}]{ILLD23}%
  \BibitemOpen
  \bibfield  {author} {\bibinfo {author} {\bibnamefont {{Institut
  Laue-Langevin}}},\ }\href@noop {} {\bibinfo {title} {Thermal neutron two-axis
  diffractometer for single-crystals {D23}}},\ \bibinfo {howpublished}
  {\url{https://www.ill.eu/users/instruments/instruments-list/d23/description/instrument-layout}}
  (\bibinfo {year} {2023})\BibitemShut {NoStop}%
\bibitem [{\citenamefont {Ressouche}\ \emph {et~al.}(2025)\citenamefont
  {Ressouche}, \citenamefont {Beauvois}, \citenamefont {Bykov}, \citenamefont
  {Ovsianikov},\ and\ \citenamefont {Usmanov}}]{Data}%
  \BibitemOpen
  \bibfield  {author} {\bibinfo {author} {\bibfnamefont {E.}~\bibnamefont
  {Ressouche}}, \bibinfo {author} {\bibfnamefont {K.}~\bibnamefont {Beauvois}},
  \bibinfo {author} {\bibfnamefont {A.}~\bibnamefont {Bykov}}, \bibinfo
  {author} {\bibfnamefont {A.}~\bibnamefont {Ovsianikov}},\ and\ \bibinfo
  {author} {\bibfnamefont {O.}~\bibnamefont {Usmanov}},\ }\href@noop {}
  {\bibinfo {title} {Datasets: Magnetic phase diagram of multiferroic and
  magnetocaloric {TmFeO3}}},\ \bibinfo {howpublished}
  {\url{https://zenodo.org/records/14997537}} (\bibinfo {year}
  {2025})\BibitemShut {NoStop}%
\bibitem [{\citenamefont {Qureshi}(2019)}]{JApplCryst2019}%
  \BibitemOpen
  \bibfield  {author} {\bibinfo {author} {\bibfnamefont {N.}~\bibnamefont
  {Qureshi}},\ }\bibfield  {title} {\bibinfo {title} {Mag2pol: A program for
  the analysis of spherical neutron polarimetry, flipping ratio and integrated
  intensity data},\ }\href {https://doi.org/10.1107/s1600576718016084}
  {\bibfield  {journal} {\bibinfo  {journal} {Journal of applied
  crystallography}\ }\textbf {\bibinfo {volume} {52}},\ \bibinfo {pages} {175}
  (\bibinfo {year} {2019})}\BibitemShut {NoStop}%
\bibitem [{Int(2006)}]{InternationalTables2006}%
  \BibitemOpen
  \href {https://doi.org/10.1107/97809553602060000100} {\bibinfo {title}
  {International tables for crystallography: Space-group symmetry}} (\bibinfo
  {year} {2006})\BibitemShut {NoStop}%
\bibitem [{\citenamefont {Bombik}\ \emph {et~al.}(2000)\citenamefont {Bombik},
  \citenamefont {Le{\'s}niewska},\ and\ \citenamefont {Pacyna}}]{Bombik2000}%
  \BibitemOpen
  \bibfield  {author} {\bibinfo {author} {\bibfnamefont {A.}~\bibnamefont
  {Bombik}}, \bibinfo {author} {\bibfnamefont {B.}~\bibnamefont
  {Le{\'s}niewska}},\ and\ \bibinfo {author} {\bibfnamefont {A.~W.}\
  \bibnamefont {Pacyna}},\ }\bibfield  {title} {\bibinfo {title} {Magnetic
  susceptibility of powder and single-crystal $\rm{TmFeO_3}$ orthoferrite},\
  }\href {https://doi.org/10.1016/s0304-8853(00)00049-4} {\bibfield  {journal}
  {\bibinfo  {journal} {Journal of Magnetism and Magnetic Materials}\ }\textbf
  {\bibinfo {volume} {214}},\ \bibinfo {pages} {243} (\bibinfo {year}
  {2000})}\BibitemShut {NoStop}%
\bibitem [{\citenamefont {Marezio}\ \emph {et~al.}(1970)\citenamefont
  {Marezio}, \citenamefont {Remeika},\ and\ \citenamefont
  {Dernier}}]{Marezio1970}%
  \BibitemOpen
  \bibfield  {author} {\bibinfo {author} {\bibfnamefont {M.}~\bibnamefont
  {Marezio}}, \bibinfo {author} {\bibfnamefont {J.}~\bibnamefont {Remeika}},\
  and\ \bibinfo {author} {\bibfnamefont {P.}~\bibnamefont {Dernier}},\
  }\bibfield  {title} {\bibinfo {title} {The crystal chemistry of the rare
  earth orthoferrites},\ }\href {https://doi.org/10.1107/s0567740870005319}
  {\bibfield  {journal} {\bibinfo  {journal} {Acta Crystallographica Section B:
  Structural Crystallography and Crystal Chemistry}\ }\textbf {\bibinfo
  {volume} {26}},\ \bibinfo {pages} {2008} (\bibinfo {year}
  {1970})}\BibitemShut {NoStop}%
\bibitem [{\citenamefont {Bombik}\ \emph {et~al.}(2001)\citenamefont {Bombik},
  \citenamefont {B{\"o}hm}, \citenamefont {Kusz},\ and\ \citenamefont
  {Pacyna}}]{Bombik2001}%
  \BibitemOpen
  \bibfield  {author} {\bibinfo {author} {\bibfnamefont {A.}~\bibnamefont
  {Bombik}}, \bibinfo {author} {\bibfnamefont {H.}~\bibnamefont {B{\"o}hm}},
  \bibinfo {author} {\bibfnamefont {J.}~\bibnamefont {Kusz}},\ and\ \bibinfo
  {author} {\bibfnamefont {A.~W.}\ \bibnamefont {Pacyna}},\ }\bibfield  {title}
  {\bibinfo {title} {Spontaneous magnetostriction and thermal expansibility of
  $\rm{TmFeO_3}$ and $\rm{LuFeO_3}$ rare earth orthoferrites},\ }\href
  {https://doi.org/10.1016/s0304-8853(01)00136-6} {\bibfield  {journal}
  {\bibinfo  {journal} {Journal of Magnetism and Magnetic Materials}\ }\textbf
  {\bibinfo {volume} {234}},\ \bibinfo {pages} {443} (\bibinfo {year}
  {2001})}\BibitemShut {NoStop}%
\bibitem [{\citenamefont {Khan}\ \emph {et~al.}(2021)\citenamefont {Khan},
  \citenamefont {Ahlawat}, \citenamefont {Deshmukh}, \citenamefont {Singh},
  \citenamefont {Sagdeo}, \citenamefont {Sathe}, \citenamefont {Karnal},\ and\
  \citenamefont {Satapathy}}]{Khan2021}%
  \BibitemOpen
  \bibfield  {author} {\bibinfo {author} {\bibfnamefont {A.~A.}\ \bibnamefont
  {Khan}}, \bibinfo {author} {\bibfnamefont {A.}~\bibnamefont {Ahlawat}},
  \bibinfo {author} {\bibfnamefont {P.}~\bibnamefont {Deshmukh}}, \bibinfo
  {author} {\bibfnamefont {M.}~\bibnamefont {Singh}}, \bibinfo {author}
  {\bibfnamefont {A.}~\bibnamefont {Sagdeo}}, \bibinfo {author} {\bibfnamefont
  {V.}~\bibnamefont {Sathe}}, \bibinfo {author} {\bibfnamefont
  {A.}~\bibnamefont {Karnal}},\ and\ \bibinfo {author} {\bibfnamefont
  {S.}~\bibnamefont {Satapathy}},\ }\bibfield  {title} {\bibinfo {title}
  {Magneto-structural correlation across the spin reorientation transition
  temperature in pure and $\rm{Sm}$ substituted $\rm{TmFeO_3}$: A temperature
  dependent raman and synchrotron \uppercase{X}-ray diffraction study},\ }\href
  {https://doi.org/10.1016/j.jallcom.2021.160985} {\bibfield  {journal}
  {\bibinfo  {journal} {Journal of Alloys and Compounds}\ }\textbf {\bibinfo
  {volume} {885}},\ \bibinfo {pages} {160985} (\bibinfo {year}
  {2021})}\BibitemShut {NoStop}%
\bibitem [{\citenamefont {Przenios{\l}o}\ \emph {et~al.}(2018)\citenamefont
  {Przenios{\l}o}, \citenamefont {Fabrykiewicz},\ and\ \citenamefont
  {Sosnowska}}]{Przenioslo2018}%
  \BibitemOpen
  \bibfield  {author} {\bibinfo {author} {\bibfnamefont {R.}~\bibnamefont
  {Przenios{\l}o}}, \bibinfo {author} {\bibfnamefont {P.}~\bibnamefont
  {Fabrykiewicz}},\ and\ \bibinfo {author} {\bibfnamefont {I.}~\bibnamefont
  {Sosnowska}},\ }\bibfield  {title} {\bibinfo {title} {Crystal symmetry
  aspects of materials with magnetic spin reorientation},\ }\href
  {https://doi.org/10.1107/s2053273318012822} {\bibfield  {journal} {\bibinfo
  {journal} {Acta Crystallographica Section A: Foundations and Advances}\
  }\textbf {\bibinfo {volume} {74}},\ \bibinfo {pages} {705} (\bibinfo {year}
  {2018})}\BibitemShut {NoStop}%
\bibitem [{\citenamefont {Tsymbal}\ \emph {et~al.}(2010)\citenamefont
  {Tsymbal}, \citenamefont {Bazaliy}, \citenamefont {Kakazei},\ and\
  \citenamefont {Vasiliev}}]{Tsymbal2010}%
  \BibitemOpen
  \bibfield  {author} {\bibinfo {author} {\bibfnamefont {L.}~\bibnamefont
  {Tsymbal}}, \bibinfo {author} {\bibfnamefont {Y.~B.}\ \bibnamefont
  {Bazaliy}}, \bibinfo {author} {\bibfnamefont {G.}~\bibnamefont {Kakazei}},\
  and\ \bibinfo {author} {\bibfnamefont {S.}~\bibnamefont {Vasiliev}},\
  }\bibfield  {title} {\bibinfo {title} {Mechanisms of magnetic and temperature
  hysteresis in $\rm{ErFeO_3}$ and $\rm{TmFeO_3}$ single crystals},\ }\bibfield
   {journal} {\bibinfo  {journal} {Journal of Applied Physics}\ }\textbf
  {\bibinfo {volume} {108}},\ \href {https://doi.org/10.1063/1.3499616}
  {10.1063/1.3499616} (\bibinfo {year} {2010})\BibitemShut {NoStop}%
\bibitem [{\citenamefont {Tsymbal}\ \emph {et~al.}(2006)\citenamefont
  {Tsymbal}, \citenamefont {Kamenev}, \citenamefont {Khara}, \citenamefont
  {Bazaliy},\ and\ \citenamefont {Wigen}}]{Tsymbal2006}%
  \BibitemOpen
  \bibfield  {author} {\bibinfo {author} {\bibfnamefont {L.}~\bibnamefont
  {Tsymbal}}, \bibinfo {author} {\bibfnamefont {V.}~\bibnamefont {Kamenev}},
  \bibinfo {author} {\bibfnamefont {D.}~\bibnamefont {Khara}}, \bibinfo
  {author} {\bibfnamefont {Y.~B.}\ \bibnamefont {Bazaliy}},\ and\ \bibinfo
  {author} {\bibfnamefont {P.}~\bibnamefont {Wigen}},\ }\bibfield  {title}
  {\bibinfo {title} {Structural properties of $\rm{TmFeO_3}$ in the spontaneous
  reorientation region},\ }\href {https://doi.org/10.1063/1.2219499} {\bibfield
   {journal} {\bibinfo  {journal} {Low Temperature Physics}\ }\textbf {\bibinfo
  {volume} {32}},\ \bibinfo {pages} {779} (\bibinfo {year} {2006})}\BibitemShut
  {NoStop}%
\bibitem [{\citenamefont {Tsymbal}\ \emph {et~al.}(2007)\citenamefont
  {Tsymbal}, \citenamefont {Bazaliy}, \citenamefont {Derkachenko},
  \citenamefont {Kamenev}, \citenamefont {Kakazei}, \citenamefont {Palomares},\
  and\ \citenamefont {Wigen}}]{Tsymbal2007}%
  \BibitemOpen
  \bibfield  {author} {\bibinfo {author} {\bibfnamefont {L.~T.}\ \bibnamefont
  {Tsymbal}}, \bibinfo {author} {\bibfnamefont {Y.~B.}\ \bibnamefont
  {Bazaliy}}, \bibinfo {author} {\bibfnamefont {V.~N.}\ \bibnamefont
  {Derkachenko}}, \bibinfo {author} {\bibfnamefont {V.~I.}\ \bibnamefont
  {Kamenev}}, \bibinfo {author} {\bibfnamefont {G.~N.}\ \bibnamefont
  {Kakazei}}, \bibinfo {author} {\bibfnamefont {F.~J.}\ \bibnamefont
  {Palomares}},\ and\ \bibinfo {author} {\bibfnamefont {P.~E.}\ \bibnamefont
  {Wigen}},\ }\bibfield  {title} {\bibinfo {title} {Magnetic and structural
  properties of spin-reorientation transitions in orthoferrites},\ }\href
  {https://doi.org/10.1063/1.2749404} {\bibfield  {journal} {\bibinfo
  {journal} {Journal of Applied Physics}\ }\textbf {\bibinfo {volume} {101}},\
  \bibinfo {pages} {123919} (\bibinfo {year} {2007})}\BibitemShut {NoStop}%
\bibitem [{\citenamefont {Fabrykiewicz}\ \emph {et~al.}(2023)\citenamefont
  {Fabrykiewicz}, \citenamefont {Przenios{\l}o},\ and\ \citenamefont
  {Sosnowska}}]{Fabrykiewicz2023}%
  \BibitemOpen
  \bibfield  {author} {\bibinfo {author} {\bibfnamefont {P.}~\bibnamefont
  {Fabrykiewicz}}, \bibinfo {author} {\bibfnamefont {R.}~\bibnamefont
  {Przenios{\l}o}},\ and\ \bibinfo {author} {\bibfnamefont {I.}~\bibnamefont
  {Sosnowska}},\ }\bibfield  {title} {\bibinfo {title} {Magnetic, electric and
  toroidal polarization modes describing the physical properties of crystals.
  $\rm{NdFeO_3}$ case},\ }\href {https://doi.org/10.1107/s2053273322009858}
  {\bibfield  {journal} {\bibinfo  {journal} {Acta Crystallographica Section A:
  Foundations and Advances}\ }\textbf {\bibinfo {volume} {79}},\ \bibinfo
  {pages} {80} (\bibinfo {year} {2023})}\BibitemShut {NoStop}%
\bibitem [{\citenamefont {Staub}\ \emph {et~al.}(2017)\citenamefont {Staub},
  \citenamefont {Rettig}, \citenamefont {Bothschafter}, \citenamefont
  {Windsor}, \citenamefont {Ramakrishnan}, \citenamefont {Avula}, \citenamefont
  {Dreiser}, \citenamefont {Piamonteze}, \citenamefont {Scagnoli},
  \citenamefont {Mukherjee} \emph {et~al.}}]{Staub2017}%
  \BibitemOpen
  \bibfield  {author} {\bibinfo {author} {\bibfnamefont {U.}~\bibnamefont
  {Staub}}, \bibinfo {author} {\bibfnamefont {L.}~\bibnamefont {Rettig}},
  \bibinfo {author} {\bibfnamefont {E.~M.}\ \bibnamefont {Bothschafter}},
  \bibinfo {author} {\bibfnamefont {Y.~W.}\ \bibnamefont {Windsor}}, \bibinfo
  {author} {\bibfnamefont {M.}~\bibnamefont {Ramakrishnan}}, \bibinfo {author}
  {\bibfnamefont {S.~R.}\ \bibnamefont {Avula}}, \bibinfo {author}
  {\bibfnamefont {J.}~\bibnamefont {Dreiser}}, \bibinfo {author} {\bibfnamefont
  {C.}~\bibnamefont {Piamonteze}}, \bibinfo {author} {\bibfnamefont
  {V.}~\bibnamefont {Scagnoli}}, \bibinfo {author} {\bibfnamefont
  {S.}~\bibnamefont {Mukherjee}}, \emph {et~al.},\ }\bibfield  {title}
  {\bibinfo {title} {Interplay of fe and tm moments through the
  spin-reorientation transition in $\rm{TmFeO_3}$},\ }\href
  {https://doi.org/10.1103/physrevb.96.174408} {\bibfield  {journal} {\bibinfo
  {journal} {Physical Review B}\ }\textbf {\bibinfo {volume} {96}},\ \bibinfo
  {pages} {174408} (\bibinfo {year} {2017})}\BibitemShut {NoStop}%
\bibitem [{\citenamefont {Zvezdin}\ \emph {et~al.}(2021)\citenamefont
  {Zvezdin}, \citenamefont {Gareeva},\ and\ \citenamefont
  {Chen}}]{Zvezdin2021}%
  \BibitemOpen
  \bibfield  {author} {\bibinfo {author} {\bibfnamefont {A.}~\bibnamefont
  {Zvezdin}}, \bibinfo {author} {\bibfnamefont {Z.}~\bibnamefont {Gareeva}},\
  and\ \bibinfo {author} {\bibfnamefont {X.}~\bibnamefont {Chen}},\ }\bibfield
  {title} {\bibinfo {title} {Multiferroic order parameters in rhombic
  antiferromagnets $\rm{RCrO_3}$},\ }\href
  {https://doi.org/10.1088/1361-648x/ac0dd6} {\bibfield  {journal} {\bibinfo
  {journal} {Journal of Physics: Condensed Matter}\ }\textbf {\bibinfo {volume}
  {33}},\ \bibinfo {pages} {385801} (\bibinfo {year} {2021})}\BibitemShut
  {NoStop}%
\bibitem [{\citenamefont {Malozemoff}\ and\ \citenamefont
  {White}(1970)}]{Malozemoff1970}%
  \BibitemOpen
  \bibfield  {author} {\bibinfo {author} {\bibfnamefont {A.}~\bibnamefont
  {Malozemoff}}\ and\ \bibinfo {author} {\bibfnamefont {R.}~\bibnamefont
  {White}},\ }\bibfield  {title} {\bibinfo {title} {Optical spectra of
  even-electron rare earth ions in the orthoferrites},\ }\href
  {https://doi.org/10.1016/0038-1098(70)90191-2} {\bibfield  {journal}
  {\bibinfo  {journal} {Solid State Communications}\ }\textbf {\bibinfo
  {volume} {8}},\ \bibinfo {pages} {665–668} (\bibinfo {year}
  {1970})}\BibitemShut {NoStop}%
\bibitem [{\citenamefont {Song}\ \emph {et~al.}(2023)\citenamefont {Song},
  \citenamefont {Fan}, \citenamefont {Jia}, \citenamefont {Sun}, \citenamefont
  {Ma}, \citenamefont {Yang}, \citenamefont {Zhu}, \citenamefont {Kang},
  \citenamefont {Feng},\ and\ \citenamefont {Cao}}]{Song2023}%
  \BibitemOpen
  \bibfield  {author} {\bibinfo {author} {\bibfnamefont {H.}~\bibnamefont
  {Song}}, \bibinfo {author} {\bibfnamefont {W.}~\bibnamefont {Fan}}, \bibinfo
  {author} {\bibfnamefont {R.}~\bibnamefont {Jia}}, \bibinfo {author}
  {\bibfnamefont {Z.}~\bibnamefont {Sun}}, \bibinfo {author} {\bibfnamefont
  {X.}~\bibnamefont {Ma}}, \bibinfo {author} {\bibfnamefont {W.}~\bibnamefont
  {Yang}}, \bibinfo {author} {\bibfnamefont {S.}~\bibnamefont {Zhu}}, \bibinfo
  {author} {\bibfnamefont {B.}~\bibnamefont {Kang}}, \bibinfo {author}
  {\bibfnamefont {Z.}~\bibnamefont {Feng}},\ and\ \bibinfo {author}
  {\bibfnamefont {S.}~\bibnamefont {Cao}},\ }\bibfield  {title} {\bibinfo
  {title} {Low field spin switching of single crystal $\rm{TmFeO_3}$},\ }\href
  {https://doi.org/10.1016/j.ceramint.2023.04.029} {\bibfield  {journal}
  {\bibinfo  {journal} {Ceramics International}\ }\textbf {\bibinfo {volume}
  {49}},\ \bibinfo {pages} {22038} (\bibinfo {year} {2023})}\BibitemShut
  {NoStop}%
\bibitem [{\citenamefont {Moskvin}\ \emph {et~al.}(2022)\citenamefont
  {Moskvin}, \citenamefont {Vasinovich},\ and\ \citenamefont
  {Shadrin}}]{Moskvin2022}%
  \BibitemOpen
  \bibfield  {author} {\bibinfo {author} {\bibfnamefont {A.}~\bibnamefont
  {Moskvin}}, \bibinfo {author} {\bibfnamefont {E.}~\bibnamefont
  {Vasinovich}},\ and\ \bibinfo {author} {\bibfnamefont {A.}~\bibnamefont
  {Shadrin}},\ }\bibfield  {title} {\bibinfo {title} {Simple realistic model of
  spin reorientation in 4\textit{f}-3\textit{d} compounds},\ }\href
  {https://doi.org/10.3390/magnetochemistry8040045} {\bibfield  {journal}
  {\bibinfo  {journal} {Magnetochemistry}\ }\textbf {\bibinfo {volume} {8}},\
  \bibinfo {pages} {45} (\bibinfo {year} {2022})}\BibitemShut {NoStop}%
\bibitem [{\citenamefont {Zhao}\ \emph {et~al.}(2014)\citenamefont {Zhao},
  \citenamefont {Zhao}, \citenamefont {Zhou}, \citenamefont {Zhang},
  \citenamefont {Li}, \citenamefont {Fan}, \citenamefont {Sun},\ and\
  \citenamefont {Li}}]{Zhao2014}%
  \BibitemOpen
  \bibfield  {author} {\bibinfo {author} {\bibfnamefont {Z.}~\bibnamefont
  {Zhao}}, \bibinfo {author} {\bibfnamefont {X.}~\bibnamefont {Zhao}}, \bibinfo
  {author} {\bibfnamefont {H.}~\bibnamefont {Zhou}}, \bibinfo {author}
  {\bibfnamefont {F.}~\bibnamefont {Zhang}}, \bibinfo {author} {\bibfnamefont
  {Q.}~\bibnamefont {Li}}, \bibinfo {author} {\bibfnamefont {C.}~\bibnamefont
  {Fan}}, \bibinfo {author} {\bibfnamefont {X.}~\bibnamefont {Sun}},\ and\
  \bibinfo {author} {\bibfnamefont {X.}~\bibnamefont {Li}},\ }\bibfield
  {title} {\bibinfo {title} {Ground state and magnetic phase transitions of
  orthoferrite $\rm{DyFeO_3}$},\ }\href
  {https://doi.org/10.1103/physrevb.89.224405} {\bibfield  {journal} {\bibinfo
  {journal} {Physical Review B}\ }\textbf {\bibinfo {volume} {89}},\ \bibinfo
  {pages} {224405} (\bibinfo {year} {2014})}\BibitemShut {NoStop}%
\bibitem [{\citenamefont {Eremenko}\ \emph {et~al.}(1987)\citenamefont
  {Eremenko}, \citenamefont {Gnatchenko}, \citenamefont {Kharchenko},
  \citenamefont {Lebedev}, \citenamefont {Piotrowski}, \citenamefont
  {Szymczak},\ and\ \citenamefont {Szymczak}}]{Eremenko1987}%
  \BibitemOpen
  \bibfield  {author} {\bibinfo {author} {\bibfnamefont {V.}~\bibnamefont
  {Eremenko}}, \bibinfo {author} {\bibfnamefont {S.}~\bibnamefont
  {Gnatchenko}}, \bibinfo {author} {\bibfnamefont {N.}~\bibnamefont
  {Kharchenko}}, \bibinfo {author} {\bibfnamefont {P.}~\bibnamefont {Lebedev}},
  \bibinfo {author} {\bibfnamefont {K.}~\bibnamefont {Piotrowski}}, \bibinfo
  {author} {\bibfnamefont {H.}~\bibnamefont {Szymczak}},\ and\ \bibinfo
  {author} {\bibfnamefont {R.}~\bibnamefont {Szymczak}},\ }\bibfield  {title}
  {\bibinfo {title} {New magnetic phase transitions in $\rm{DyFeO_3}$},\
  }\href@noop {} {\bibfield  {journal} {\bibinfo  {journal} {Europhysics
  Letters}\ }\textbf {\bibinfo {volume} {4}},\ \bibinfo {pages} {1327}
  (\bibinfo {year} {1987})}\BibitemShut {NoStop}%
\bibitem [{\citenamefont {Guo}\ \emph {et~al.}(2020)\citenamefont {Guo},
  \citenamefont {Cheng}, \citenamefont {Ren}, \citenamefont {Zhang},
  \citenamefont {Lin}, \citenamefont {Jin}, \citenamefont {Cao}, \citenamefont
  {Sheng},\ and\ \citenamefont {Ma}}]{Guo2020}%
  \BibitemOpen
  \bibfield  {author} {\bibinfo {author} {\bibfnamefont {J.}~\bibnamefont
  {Guo}}, \bibinfo {author} {\bibfnamefont {L.}~\bibnamefont {Cheng}}, \bibinfo
  {author} {\bibfnamefont {Z.}~\bibnamefont {Ren}}, \bibinfo {author}
  {\bibfnamefont {W.}~\bibnamefont {Zhang}}, \bibinfo {author} {\bibfnamefont
  {X.}~\bibnamefont {Lin}}, \bibinfo {author} {\bibfnamefont {Z.}~\bibnamefont
  {Jin}}, \bibinfo {author} {\bibfnamefont {S.}~\bibnamefont {Cao}}, \bibinfo
  {author} {\bibfnamefont {Z.}~\bibnamefont {Sheng}},\ and\ \bibinfo {author}
  {\bibfnamefont {G.}~\bibnamefont {Ma}},\ }\bibfield  {title} {\bibinfo
  {title} {Magnetic field tuning of spin resonance in $\rm{TmFeO_3}$ single
  crystal probed with \uppercase{TH}z transient},\ }\href
  {https://doi.org/10.1088/1361-648x/ab6d0f} {\bibfield  {journal} {\bibinfo
  {journal} {Journal of Physics: Condensed Matter}\ }\textbf {\bibinfo {volume}
  {32}},\ \bibinfo {pages} {185401} (\bibinfo {year} {2020})}\BibitemShut
  {NoStop}%
\bibitem [{\citenamefont {Wang}\ \emph {et~al.}(2023)\citenamefont {Wang},
  \citenamefont {Zhu}, \citenamefont {Hu}, \citenamefont {Ju}, \citenamefont
  {Su}, \citenamefont {Huang}, \citenamefont {Chen}, \citenamefont {Cao},\ and\
  \citenamefont {Wang}}]{Wang2023}%
  \BibitemOpen
  \bibfield  {author} {\bibinfo {author} {\bibfnamefont {N.}~\bibnamefont
  {Wang}}, \bibinfo {author} {\bibfnamefont {G.}~\bibnamefont {Zhu}}, \bibinfo
  {author} {\bibfnamefont {Z.}~\bibnamefont {Hu}}, \bibinfo {author}
  {\bibfnamefont {X.}~\bibnamefont {Ju}}, \bibinfo {author} {\bibfnamefont
  {H.}~\bibnamefont {Su}}, \bibinfo {author} {\bibfnamefont {F.}~\bibnamefont
  {Huang}}, \bibinfo {author} {\bibfnamefont {Q.}~\bibnamefont {Chen}},
  \bibinfo {author} {\bibfnamefont {Y.}~\bibnamefont {Cao}},\ and\ \bibinfo
  {author} {\bibfnamefont {X.}~\bibnamefont {Wang}},\ }\bibfield  {title}
  {\bibinfo {title} {Magnetic properties of rare-earth orthoferrites},\ }\href
  {https://doi.org/10.1016/j.infrared.2023.104937} {\bibfield  {journal}
  {\bibinfo  {journal} {Infrared Phys. Technol.}\ }\textbf {\bibinfo {volume}
  {135}},\ \bibinfo {pages} {104937} (\bibinfo {year} {2023})}\BibitemShut
  {NoStop}%
\bibitem [{\citenamefont {Belov}\ \emph {et~al.}(1979)\citenamefont {Belov},
  \citenamefont {Zvezdin},\ and\ \citenamefont {Mukhin}}]{Belov1979}%
  \BibitemOpen
  \bibfield  {author} {\bibinfo {author} {\bibfnamefont {K.}~\bibnamefont
  {Belov}}, \bibinfo {author} {\bibfnamefont {A.}~\bibnamefont {Zvezdin}},\
  and\ \bibinfo {author} {\bibfnamefont {A.}~\bibnamefont {Mukhin}},\
  }\bibfield  {title} {\bibinfo {title} {Magnetic phase transformations in
  \uppercase{T}b orthoferrites},\ }\href@noop {} {\bibfield  {journal}
  {\bibinfo  {journal} {Zhurnal Eksperimental'noi i Teoreticheskoi Fiziki}\
  }\textbf {\bibinfo {volume} {76}},\ \bibinfo {pages} {1100} (\bibinfo {year}
  {1979})}\BibitemShut {NoStop}%
\end{thebibliography}%

\end{document}